%% file: main.tex
\documentclass[a4paper,11pt]{article}
\pdfoutput=1

\usepackage{graphicx}
\usepackage{amssymb}
\usepackage{subcaption}

\usepackage{color}

\usepackage[colorinlistoftodos]{todonotes}

\usepackage{verbatim}

\usepackage{tikz}

\usepackage[perpage]{footmisc}

\usepackage{lineno}

\usepackage{booktabs}

\usepackage{listings}

\usepackage{jinstpub} % for details on the use of the package, please
\title{Design and performance of a 35-ton liquid argon time projection chamber as a prototype for future very large detectors}

\input TPC_authorlist2.txt
\input TPC_instlist2.txt

\emailAdd{M. Convery (convery@slac.stanford.edu)}
\emailAdd{T. Junk (trj@fnal.gov)}

\input abstract.tex


\begin{document}
\maketitle
\flushbottom

%% main text
\input introduction_section.tex

\input detectordesign_section.tex

\input trigger_section.tex

\input runningconditions_section.tex

\input rawdata_section.tex

\input dataprocessing_section.tex

\input hitfinding_section.tex

\input relativealignment_section.tex

\input zgap_section.tex

\input apacrosser_section.tex

\input lifetime_section_v5.tex

\input diffusion_section.tex

\input summary_section.tex

%\begin{figure}
%  \centering
%  \includegraphics[width=8cm]{evd_pizero_candidate.png}
%  \caption[Event display showing two electromagnetic showers and two tracks.]
%          {Event display showing two electromagnetic showers and two tracks.
%   The start of the readout window is later than the interaction time as can be seen
%    by the crossing of one of the showers and the APA plane (blue dashed line). % The two
%    electromagnetic showers may have been caused by the decay of a $\pi^0$.}
%  \label{fig:evd_pizero_candidate}
% \end{figure}

\clearpage
\newpage

\input bibliography.tex

%% The Appendices part is started with the command \appendix;
%% appendix sections are then done as normal sections
%% \appendix

%% \section{}
%% \label{}

%% References
%%
%% Following citation commands can be used in the body text:
%% Usage of \cite is as follows:
%%   \cite{key}          ==>>  [#]
%%   \cite[chap. 2]{key} ==>>  [#, chap. 2]
%%   \citet{key}         ==>>  Author [#]

%% References with bibTeX database:
% not yet provided

%% Authors are advised to submit their bibtex database files. They are
%% requested to list a bibtex style file in the manuscript if they do
%% not want to use model1-num-names.bst.

%% References without bibTeX database:

% \begin{thebibliography}{00}

%% \bibitem must have the following form:
%%   \bibitem{key}...
%%

% \bibitem{}

% \end{thebibliography}

\end{document}

%% file: TPC_authorlist2.txt
\author[3]{D.~L.~Adams,}
\author[26]{M.~Baird \footnote{Now at the University of Virginia},}
\author[19]{G.~Barr,}
\author[20]{N.~Barros \footnote{Now at Laborat\'{o}rio de Instrumenta\c{c}\~{a}o e F\'{i}sica Experimental de Part\'{i}culas, Lisbon, Portugal
},}
\author[15]{A.~Blake,}
\author[17]{E.~Blaufuss,}
\author[26]{A.~Booth,}
\author[15]{D. ~Brailsford,}
\author[6]{N.~Buchanan,}
\author[11]{B.~Carls \footnote{Now at Commonwealth Edison},}
\author[3]{H.~Chen,}
\author[23]{M.~Convery,}
\author[3]{G.~De Geronimo,}
\author[15]{T.~Dealtry,}
\author[2]{R.~Dharmapalan\footnote{Now at the University of Hawaii},}
\author[2]{Z.~Djurcic,}
\author[7]{J.~Fowler,}
\author[20]{S.~Glavin,}
\author[10]{R.~A.~Gomes,}
\author[2]{M.~C.~Goodman,}
\author[23]{M.~Graham,}
\author[27]{L.~Greenler,}
\author[11]{A.~Hahn,}
\author[26]{J.~Hartnell,}
\author[23]{R.~Herbst,}
\author[13]{A.~Higuera,}
\author[11]{A.~Himmel,}
\author[16]{J.~Insler\footnote{Now at Slater Matsil, LLP},}
\author[17]{J.~Jacobsen,}
\author[11]{T.~Junk,}
\author[3]{B.~Kirby,}
\author[20]{J.~Klein,}
\author[22]{V.~A.~Kudryavtsev,}
\author[16]{T.~Kutter,}
\author[3]{Y.~Li,}
\author[25]{X.~Li,}
\author[6]{S.~Lin,}
\author[22]{N.~McConkey\footnote{Now at the University of Manchester},}
\author[9]{C.~A.~Moura,}
\author[14]{S.~Mufson,}
\author[3]{N.~Nambiar\footnote{Now at Teradyne Inc.},}
\author[15]{J.~Nowak,}
\author[8]{M.~Nunes,}
\author[27]{R.~Paulos,}
\author[3]{X.~Qian,}
\author[8]{O. ~Rodrigues \footnote{Now at Syracuse University},}
\author[21]{W.~Sands,}
\author[25]{G.~Santucci,}
\author[3]{R.~Sharma,}
\author[7]{G.~Sinev,}
\author[22]{N.~J.~C.~Spooner,}
\author[1]{I.~Stancu,}
\author[5]{D.~Stefan \footnote{Now at RnD Team Design Studio},}
\author[3]{J.~Stewart,}
\author[24]{J.~Stock,}
\author[11]{T.~Strauss,}
\author[18]{R.~Sulej  \footnote{Now at RnD Team Design Studio},}
\author[12]{Y.~Sun \footnote{Now at Fermi National Accelerator Laboratory},}
\author[22]{M.~Thiesse,}
\author[22]{L.~F.~Thompson,}
\author[23]{Y.~T.~Tsai,}
\author[20]{R.~Van Berg,}
\author[8]{T.~Vieira,}
\author[22]{M.~Wallbank \footnote{Now at the University of Cincinnati},}
\author[4]{H.~Wang,}
\author[4]{Y.~Wang,}
\author[22]{T.~K.~Warburton \footnote{Now at Iowa State University},}
\author[27]{D.~Wenman,}
\author[14]{D.~Whittington\footnote{Now at Syracuse University},}
\author[6]{R.~J.~Wilson,}
\author[3]{M.~Worcester,}
\author[11]{T.~Yang,}
\author[3]{B.~Yu,}
\author[3]{and C.~Zhang}

%% file: TPC_instlist2.txt
\affiliation[1]{University of Alabama, Tuscaloosa, AL 35487, USA}		
\affiliation[2]{Argonne National Laboratory, Argonne, IL 60439, USA}		
\affiliation[3]{Brookhaven National Laboratory, Upton, NY 11973, USA}		
\affiliation[4]{University of California Los Angeles, Los Angeles, CA 90095, USA}		
\affiliation[5]{CERN, European Organization for Nuclear Research 1211 Geneve 23, Switzerland, CERN}		
\affiliation[6]{Colorado State University, Fort Collins, CO 80523, USA}		
\affiliation[7]{Duke University, Durham, NC 27708, USA}		
\affiliation[8]{Universidade Estadual de Campinas, Campinas - SP, 13083-970, Brazil}		
\affiliation[9]{Universidade Federal do ABC, Santo Andr\'e - SP, 09210-580, Brazil}		
\affiliation[10]{Universidade Federal de Goias, Instituto de Fisica, Goiania, GO 74690-900, Brazil}		
\affiliation[11]{Fermi National Accelerator Laboratory, Batavia, IL 60510, USA}		
\affiliation[12]{University of Hawaii, Honolulu, HI 96822, USA}		
\affiliation[13]{University of Houston, Houston, TX 77204, USA}		
\affiliation[14]{Indiana University, Bloomington, IN 47405, USA}		
\affiliation[15]{Lancaster University, Bailrigg, Lancaster LA1 4YB, United Kingdom}		
\affiliation[16]{Louisiana State University, Baton Rouge, LA 70803, USA}		
\affiliation[17]{University of Maryland, College Park, MD 20742, USA}		
\affiliation[18]{National Centre for Nuclear Research, A. Soltana 7, 05 400 Otwock, Poland}		
\affiliation[19]{University of Oxford, Oxford, OX1 3RH, United Kingdom}		
\affiliation[20]{University of Pennsylvania, Philadelphia, PA 19104, USA}		
\affiliation[21]{Princeton University, Princeton, NJ 08544, USA}		
\affiliation[22]{University of Sheffield, Department of Physics and Astronomy, Sheffield S3 7RH, United Kingdom}		
\affiliation[23]{SLAC National Accelerator Laboratory, Menlo Park, CA 94025, USA}		
\affiliation[24]{South Dakota School of Mines and Technology, Rapid City, SD 57701, USA}		
\affiliation[25]{Stony Brook University,Stony Brook, New York 11794, USA}		
\affiliation[26]{University of Sussex, Brighton, BN1 9RH, United Kingdom}		
\affiliation[27]{University of Wisconsin (Madison), Madison, WI 53706, USA}		

%% file: abstract.tex
\abstract{
Liquid argon time projection chamber technology is an attractive
choice for large neutrino detectors, as it provides a high-resolution
active target and it is expected to be scalable to very large masses.
Consequently, it has been chosen as the technology for the first
module of the DUNE far detector.  However, the fiducial mass required
for "far detectors" of the next generation of neutrino oscillation
experiments far exceeds what has been demonstrated so far.  Scaling to
this larger mass, as well as the requirement for underground
construction places a number of additional constraints on the design.
A prototype 35-ton cryostat was built at Fermi National Acccelerator
Laboratory to test the functionality of the components foreseen to be
used in a very large far detector.  The Phase~I run, completed in
early 2014, demonstrated that liquid argon could be maintained at
sufficient purity in a membrane cryostat.  A time projection chamber
was installed for the Phase~II run, which collected data in February
and March of 2016.  The Phase~II run was a test of the modular anode
plane assemblies with wrapped wires, cold readout electronics, and
integrated photon detection systems.  While the details of the design
do not match exactly those chosen for the DUNE far detector, the
35-ton TPC prototype is a demonstration of the functionality of the
basic components.  Measurements are performed using the Phase~II data
to extract signal and noise characteristics and to align the detector
components.  A measurement of the electron lifetime is presented, and
a novel technique for measuring a track's position based on pulse
properties is described.  
}

%% file: introduction_section.tex
\section{Introduction}

The single-phase liquid argon time projection chamber (LArTPC) has been
demonstrated to be an effective neutrino detector technology in 
ICARUS~\cite{icarusnim} and MicroBooNE~\cite{microboone}.
However, scaling this technology to the fiducial mass required for the next generation
of long-baseline experiments requires modification
of several design elements.
Furthermore, locating a large LArTPC deep underground
places further requirements on the design. 
The Long-Baseline Neturino Experiment (LBNE) Collaboration proposed
a LArTPC design to address these requirements~\cite{LBNE}.
When the Deep Underground Neutrino Experiment (DUNE) Collaboration 
was formed and superseded the LBNE effort, it adopted many of the same ideas for its 
far detector (FD)~\cite{dunecdrvol4,dunetdrvol1,dunetdrvol4}. 

The 35-ton prototype was designed to test the performance of the
concepts and components proposed by LBNE and largely adopted by DUNE. 
The DUNE FD is proposed to
consist of 40~ktons (fiducial) of liquid argon in four 10~kton modules
located at the 4850'~level of the Sanford Underground Research
Facility (SURF)~\cite{surf} in Lead, South Dakota.  The start of
installation of the first 10~kton module is scheduled to begin in
2022.  The first DUNE FD module is planned to be a single-phase LArTPC.
Subsequent modules may be additional single-phase modules or
dual-phase modules~\cite{dualphase1,dualphase2,dualphase3,dualphase4}.

The DUNE FD modules will be much larger than any previous
LArTPC.  The components must be shipped
to the site, lowered down the shaft, assembled in place, tested, and
operated, all in a cost-effective and time-efficient manner.  These
steps place constraints on the design of the FD, and
compromises must be made in order to satisfy these constraints.  To
meet the physics goals of DUNE, the performance of the detector must
satisfy basic requirements of spatial, time, and energy resolution,
signal-to-noise (S/N) performance, detection efficiency and uptime.  The
design choices must be tested in prototypes before the FD
design is finalised and resources are committed.
The first phase of the 35-ton prototype's operation, which was conducted
without a time projection chamber (TPC) installed, demonstrated that
the required electron lifetime is achievable in a 
non-evacuated membrane cryostat~\cite{Hahn:2016tia,Montanari:2015zwa}.
This paper focuses on the TPC aspects of the 35-ton prototype. 
A previous paper~\cite{Adams:2018lfb} focuses on its photon detection system.
Section~\ref{sec:detector} describes the design of the 35-ton
prototype and which design choices for the FD are tested.
The trigger system is described in section~\ref{sec:trigger}.  The
data acquisition system is described in section~\ref{sec:daq}, and the
running conditions are summarised in section~\ref{sec:run}.  Several
analyses of the data from the Phase~II run of the 35-ton prototype are
listed in sections~\ref{sec:dataproc} through~\ref{sec:diffusion}.
These comprise studies of the signal and noise performance of the
system, the relative alignment of the external counters and the TPC
using cosmic-ray tracks, the measurement of the relative time between
the external counters and the TPC using tracks that cross the
anode-plane assembly (APA) volumes, alignment and charge
characteristic measurements using tracks that cross between one APA's
drift volume to another's, a measurement of the electron lifetime, and
studies of diffusion of drifting electrons.  A summary and outlook is
given in section~\ref{sec:summary}.

Because of the rapid evolution of DUNE's FD design, the choices
considered when the 35-ton prototype design was finalised are no
longer exactly those considered, although
the broad features are the same. Section~\ref{sec:detector} describes
these issues in detail. 
Furthermore, the analyses presented here use early versions 
of the simulation and reconstruction software, and newer variations on the noise-reduction
techniques, such as those described in~\cite{Acciarri:2017sde}, are not applied.
Subsequently, the ProtoDUNE Single Phase prototype (ProtoDUNE-SP),
which has a design closer to that now planned for the FD,
was constructed and operated at CERN in late 2018~\cite{protodunesp}.
ProtoDUNE-SP benefits from lower noise operation and more sophisticated analysis techniques.
The 35-ton prototype and its data analysis are the first attempts at a ``DUNE-style'' 
LArTPC and provide key insights to the more advanced hardware designs and software.

%% file: detectordesign_section.tex
\section{Detector design}
\label{sec:detector}

\begin{figure}[ht]
\begin{center}
{\includegraphics[width=0.8\textwidth]{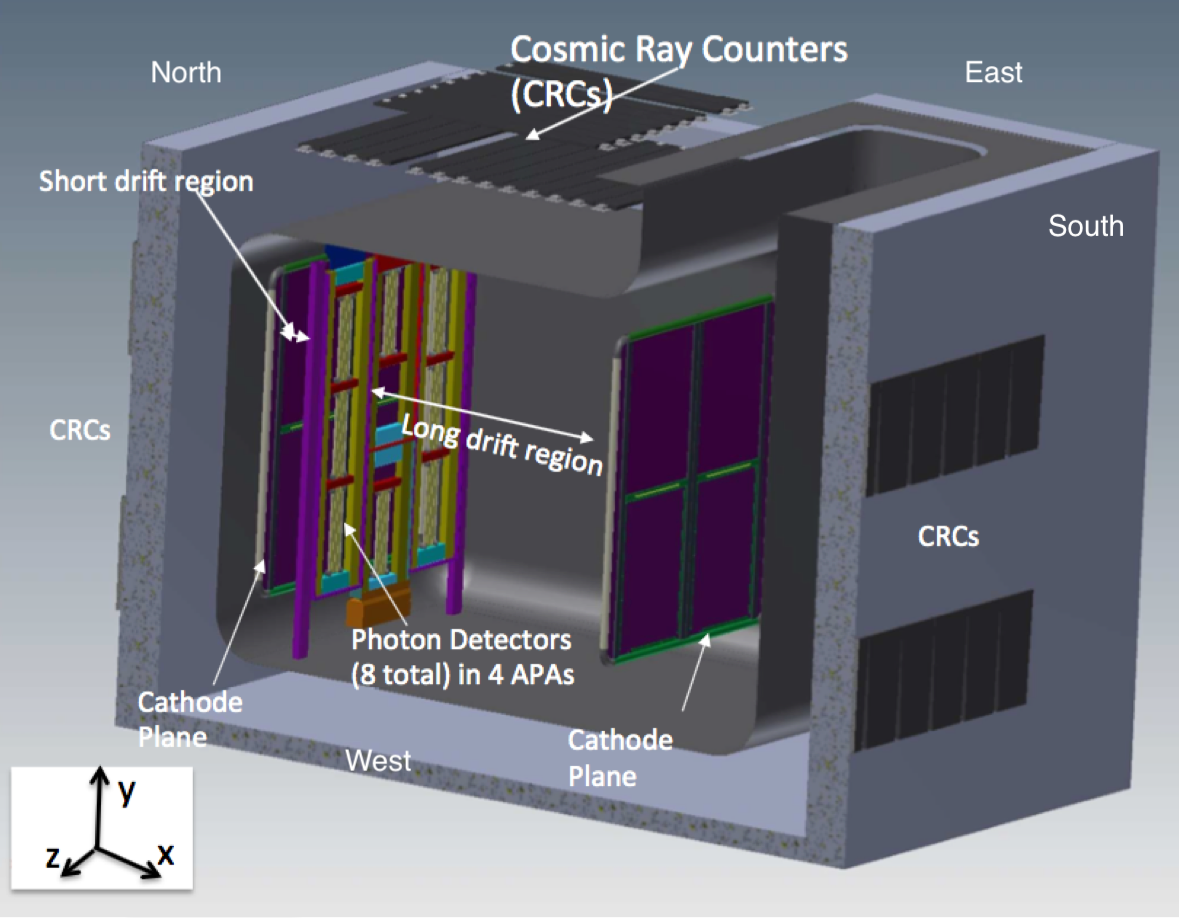}};
\end{center}
\caption{\label{fig:tpcdrawing} Drawing of the TPC within the
  cryostat. Critical components are labeled and the coordinate system is
  defined.}
\end{figure}

\begin{figure}[ht]
\includegraphics[width=0.45\textwidth]{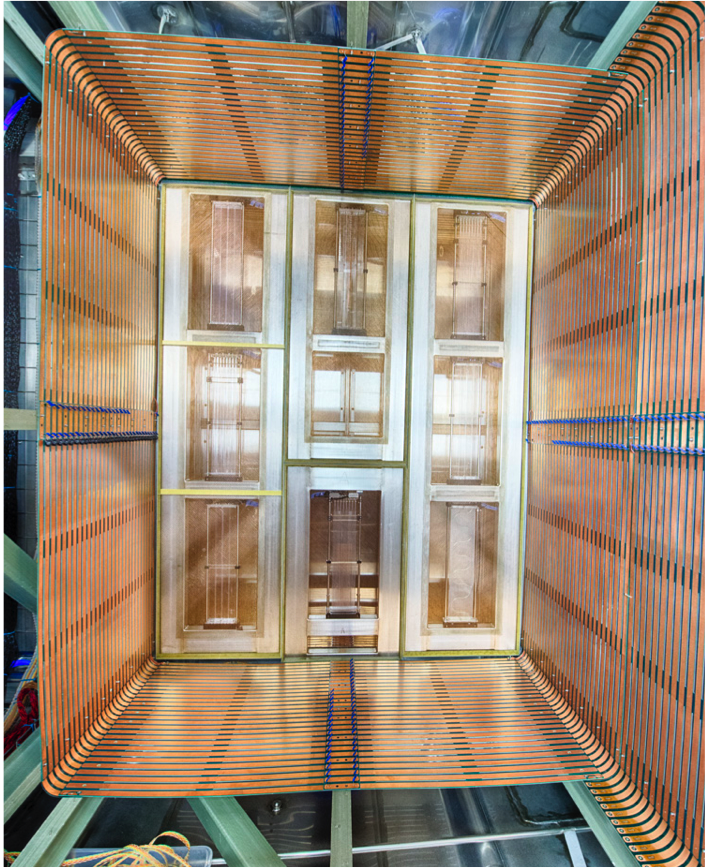}\quad \includegraphics[width=0.49\textwidth]{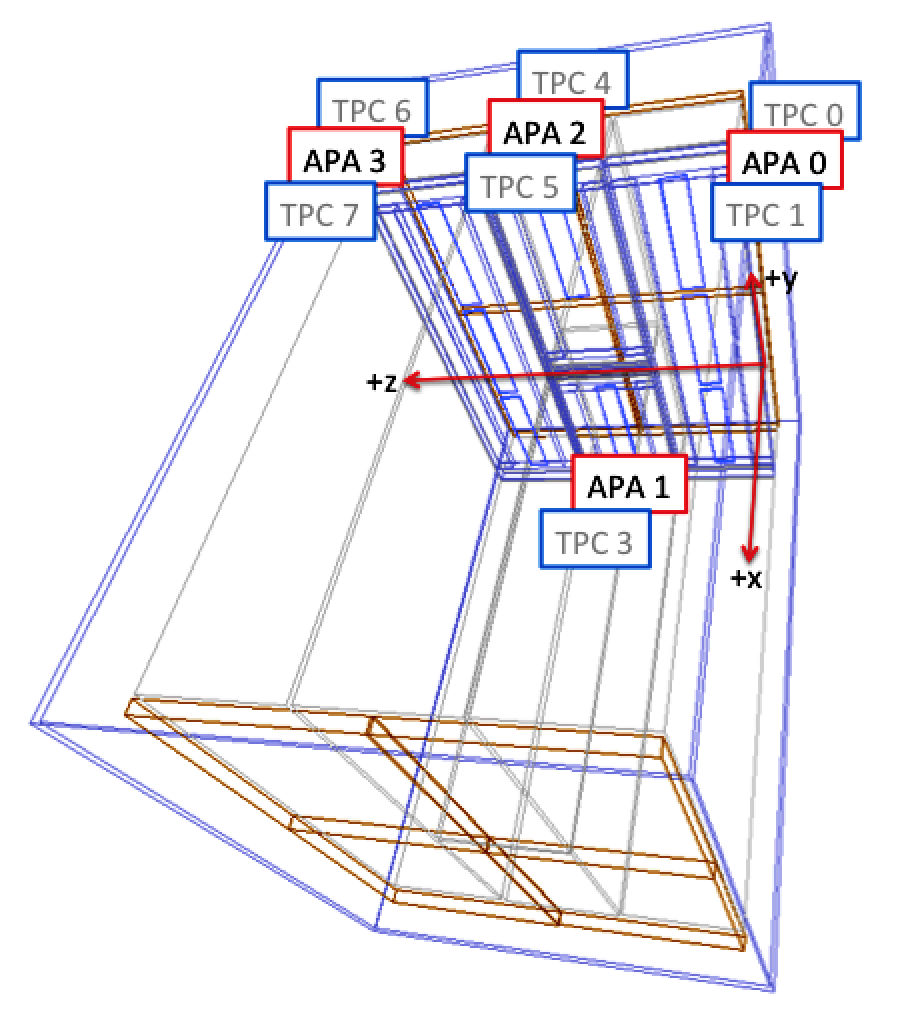}
\caption{\label{fig:tpcphotonumbering} (left) A photograph taken inside the cryostat
  during construction. The APA plane and partially constructed field
  cage is visible. (right)   Numbering of the drift volumes (TPCs) in the 35-ton prototype.  Even-numbered
  TPCs are in the short drift volume and odd-numbered TPCs are in the long drift volume.
  TPC number 2 is not labeled and is behind TPC number 3.}
\end{figure}

%The 35-ton prototype detector was built in order to test the design of
%the LBNE FD~\cite{LBNE}, a very large LArTPC to be located at
%the 4850 foot level at SURF.  This design has evolved over time, but its main
%features have been adopted for the proposed DUNE FD~\cite{dunecdrvol4}.
The critical design choices for the DUNE FD are described below, as well as the
elements of the 35-ton prototype's design that test these choices.
Figures \ref{fig:tpcdrawing} and
\ref{fig:tpcphotonumbering} show a drawing of the 35-ton TPC within its
cryostat, a photograph of the TPC interior, and the numbering scheme for the drift volumes. 

Instead of using a single frame holding the anode wires, which has been typical of previous
LArTPCs, the DUNE FD's anode planes will comprise many APAs.
In order to ship the APAs from their manufacturing site to SURF in
standard high-cube shipping containers, lower them down the shaft at
SURF and install them in the cryostat, they are limited in size to
6.3~m~$\times$~2.3~m.  Amplifiers and digitisers are placed in the
cryostat in order to reduce thermal noise and to simplify the cabling.
The FD has two layers of APAs stacked vertically.  The
electronics are mounted on the bottom of the bottom layer and on the top of the
top layer.  In order to minimise the effects of
electron lifetime and diffusion, as well as to reduce the required high voltage
(HV), the drift length in the DUNE FD is limited to~3.6~m.  This requires the APAs to
be placed within the active volume and to be read out on both sides.
Each side has its own plane of vertical collection, ``$Z$'', wires, but
the induction wires are wrapped around the APA and are thus shared
between the two sides.  There are two induction planes ($U$ and $V$),
the wires of which are at angles relative to collection plane and wrap
around the APA and thus measure signals on both sides of the APA.  For
the DUNE FD, these angles have been chosen to be approximately
37$^\circ$.   On each side of an APA, an uninstrumented grid wire plane
is situated between the $U$ plane and the drift volume, and a grounded
mesh is installed between the collection plane and the argon volume
in the middle of the APA frame where the photon detectors lie.  The $V$ wires are held at ground,
as is the mesh.  The potentials of the grid, $U$, and $Z$ wires are chosen so that
all wire planes are transparent to drifting charge except the collection plane,
which has a high efficiency for collecting drifting charge.

The 35-ton prototype was designed to test the performance of a detector
with these choices.  However, in order to fit inside the membrane
cryostat of the Phase~I prototype, some differences were necessary.
The APAs and the drift volumes were shortened relative to the
FD design.  A drift region as long as possible to fit in the
35-ton cryostat was designed, while still having a shorter drift
region on the other side of the plane containing the APAs in order to
test the double-sided readout functionality of the APAs.  The long
drift length of the 35-ton prototype is 2.225~m from the collection-plane wires in the APAs to the cathode, while the short drift length is 0.272~m.  The
induction wire angles are $45.705^\circ$ ($U$ wires) and $-44.274^\circ$ ($V$ wires)
with respect to the collection-plane wires.  The small difference in angles
is designed to aid in resolving ambiguities.  In the long APAs, each 
induction-plane wire wraps twice around the APA frame.
Each of the four APAs contains 144 $U$-plane wires, 144 $V$-plane wires,
and 224 collection-plane wires, 112 of which are on each side. 
At a temperature of 88~K, the nominal intra-plane wire spacing was chosen to be:
4.878~mm for the $U$~plane, 5.001~mm for the $V$~plane and 4.490~mm
for the collection plane, and the inter-plane spacing was chose to be 4.730~mm.

Figures \ref{fig:tpcdrawing} and \ref{fig:tpcphotonumbering} show the four APAs
in the TPC: two tall ones (APAs~0 and~3) on either side of a stack of two shorter
ones (APAs~1 and~2).  This arrangement allows the study of the gap region between APAs.  The two
tall APAs measure 2.0~m vertically by 0.5~m horizontally, and extend
from the bottom of the detector to the top.  Their electronics are
mounted on the top.  Two shorter APAs are mounted between the two long
ones, both 0.5~m wide.  The short APA on top is 1.2~m tall while the
short APA on the bottom is 0.91~m tall.  The electronics for the short
APA on the bottom are mounted on its bottom edge.  The layout of the
APAs is designed so that there are horizontal and vertical gaps
between the APAs, as there are in the DUNE FD.  The aspect
ratio of the APA frames in the 35-ton prototype is narrower than the
$2.3~{\rm m} \times 6.0~{\rm m}$ DUNE FD APA design. The 35-ton APA frame
dimensions were chosen so that they would fit in an access hatch on
the top of the cryostat.

The short middle APA (APA~1 in figure~\ref{fig:tpcdrawing}) in the 35-ton prototype was built without the
grounded meshes between the collection planes in order to test the
impact on operations and measurements.  Installed in the vertical gap
between the short middle APA and one of the long APAs is an
electrostatic deflector, which is designed to control the electric
field in this difficult-to-model region and make the charge collection
on the neighbouring wires easier to understand.  The effect on the charge
measurements as functions of bias voltage on the deflector was not
studied however.

Photon detector modules~\cite{Adams:2018lfb} are installed
between the grounded meshes of each APA, and between the collection 
planes of APA~1.
There are three designs for the
light collectors: acrylic bars coated with wavelength-shifting tetraphenyl butadiene (TPB),
acrylic fibers coated with TPB, 
and acrylic bars with wavelength-shifting fibers embedded in them.  The light from each collector
is detected by a set of silicon photomultipliers.  The signals are amplified, digitised, and recorded
as functions of time along with the TPC wire data.  The photon detector signals provide accurate
timing information for activity in the TPC, which is important for determining the absolute distance
between the charge deposition point and the anode plane.

A Cartesian coordinate system is used throughout this article.  The coordinate system is shown in figure~\ref{fig:tpcdrawing} along with the locations of the detector components.  The
$x$~axis points along the electric field, perpendicular to the APA
frames, opposite to the direction of electron drift in the long drift
volume.  In this article, ``south'' is the direction along the
positive $x$~axis.  The $y$~axis is vertical, pointing upwards, and
the horizontal $z$ axis, which points west, completes a right-handed
coordinate system.  The APA frames are in the $yz$ plane, and the
collection wires run along the $y$~axis.  The collection wires are
called $Z$ wires because they differ from each other in their $z$
coordinate and thus measure $z$.  

The 35-ton prototype detector is not in a test beam.  Cosmic rays
provide the particles required to understand its performance.  In
order to trigger on cosmic rays that provide the most information
about the detector, cosmic-ray counters (CRCs) consisting of
scintillator paddles are installed on the four
vertical walls of the steel-reinforced concrete structure supporting the cryostat.  The
scintillation light from each CRC is detected by a photomultiplier
tube (PMT).  
The analog signals from the PMTs are amplified and discriminated with
a custom circuit located 2~cm from the PMT base~\cite{counters}.
The signals are used for triggering and saved to the datastream as described 
in section~\ref{sec:trigger}.

The CRC paddles were formerly part of the CDF muon upgrade
detectors~\cite{Artikov:2004ew}.  
Each black trapezoid on the cryostat wall in figure~\ref{fig:tpcdrawing}
represents a pair of counters and measures 24.8~inches (63~cm) high, 10.7~inches (27.2~cm) wide on the narrow side, and 12.8 inches (32.5~cm) wide on the wider side.  The counter pairs were installed on the cryostat
walls in an alternating pattern to minimise dead space between
adjacent counters.

Figure~\ref{fig:tpcdrawing} shows the locations of the CRCs.  The
north and south cryostat walls each have two horizontal rows of counters, extending along the $z$ direction.
Each row consists of six counter pairs.  The pairs of counters are stacked along
$x$ for purposes of forming coincindence triggers. The east and west walls, which are not visible in the drawing,
have only one row of counters each due to obstructions present in the
experimental hall.  These rows each consist of ten counter pairs arranged along the $x$ direction.  The two counters in a counter pair are stacked along $z$.  The counters on the west wall are located near the top of the wall while those
on the east wall are located near the bottom in order to cover the active TPC volume and increase the rate of
coincidences above the horizontal-muon rate.  The heights are chosen so that a muon traversing from an upper
row on one wall and a lower row on the opposite wall will traverse the active volume of the TPC from the upper edge on one side to the lower edge of the other side.
An additional four counter-pair stacks are installed on the east wall below the bottom row in order to get improved coverage
of APA~1's volume.  There is also a set of CRCs located above
the detector, which form a muon telescope.
However, these are not used for the measurements discussed in
this article.

%\begin{figure}[ht]
%\begin{center}
%\includegraphics[width=0.7\textwidth]{annotate_counterlocs.pdf}
%\end{center}
%\caption{\label{fig:mycounters} The layout of the cosmic-ray counters used
%  for triggering.  The muon telescope counters located far above the
%  detector are not shown.}
%\end{figure}

The cryogenic system, including the cooling, purification and
monitoring systems, was adapted from that used by the Liquid Argon
Purity Demonstrator~\cite{lapd}.

Four purity monitors were installed on a vertical support in the
liquid argon, outside of the TPC volume.  In each one, ultraviolet light from a xenon flashlamp 
illuminates a cathode which emits electrons that drift through a
short drift volume and are collected by an anode.  Comparison of the
integrated charge collected in short pulses between that emitted by
the cathode and collected by the anode provided four measurements of
the electron lifetime.  Electrons that attach to impurities drift with
much smaller velocities and do not contribute to the short-pulse
charge integration.

The cathode planes were constructed out of stainless steel sheets with reinforcing bars installed
midway through in order to maintain the necessary stiffness and minimise distortions.  The voltages
were provided by a high-precision Heinzinger HV supply with a maximum output voltage of 150~kV.
High voltage was supplied to the cathode via a feedthrough
which made contact with a cup mounted on the cathode frame.  Resistors totaling
1.2~G$\Omega$ were installed in series with the high-voltage supply in order to
reduce ripple and limit the speed of charging and discharge.   % LBNE-doc-9153-v2

A set of eight CMOS CCD cameras~\cite{McConkey:2016spe,Warburton:2017ixr,Thiesse:2017qom,Wallbank:2018otb} were installed
to monitor the cryostat for potential HV breakdowns and to monitor the operations of cryogenic
components.  They viewed the argon volume between
the cathode on the long-drift side of the TPC and the cryostat, as well as the ullage
and the volume near the HV feedthrough.   Cameras were also installed to monitor the
cooldown sprayers and the phase separator, in order verify proper operation.

Low-voltage electrical power to the detector elements and signals from the FEMBs, the photon detectors and the cameras pass through a custom board called the flange board, which penetrates a flange on the top of the cryostat.

%% file: trigger_section.tex
\section{Trigger}
\label{sec:trigger}
A custom set of electronics is used to trigger the detector, to
provide timestamps to triggered events, and to provide calibration
signals to some of the subsystems. The hardware for the trigger
comprises a front end that receives and translates signals from the
counters and other subsystems, and a MicroZed evaluation board
carrying a Xilinx Zynq 7020 system-on-a-chip, that includes both an
extensive FPGA and an embedded ARM core processor running Linux.

The trigger board receives 146 digital signals from the CRCs: 96 from the side-wall counters, and 50 from the telescope.
% see LBNE-doc-10184-v2
For triggers based on the CRC, the two signals 
from each trapezoidal counter pair are logically ANDed on the trigger board to reduce
accidentals.
These signals are then compared to a programmable trigger mask.  Hardware trigger signals generated in the FPGA
are sent to all subsystems, including the downstream DAQ readout, and
information regarding which trigger had occurred and its timestamp 
are also sent to the event builder.

Given the speed of the Zynq~7020 and the high bandwidth of the
Ethernet connection available on the MicroZed, the times of all
counter hits are also streamed continuously, so that offline
triggering is also possible.

For the analyses presented in this paper, pairs of CRCs (East/West or
North/South) are used to trigger the events.  Each event thus comes
tagged with an event time ($t_0$) and a rough measure of its track
direction and position, which provides a useful set of tracks for
evaluating the performance of the 35-ton prototype detector.

\section{Data acquisition}
\label{sec:daq}

The currents on the wires were amplified by cold preamplifiers and
digitised by 12-bit ADCs, also in the cold volume.  Front-end ASICs~\cite{larasic}
contain the preamplifiers for sixteen channels apiece. 
The front-end
ASICs allow for the remote configuration of the preamplifier settings.
There are four gain settings: 4.7, 7.8, 14, 25~mV$\!$/fC, and four
shaping-time settings: 0.5~$\mu$s, 1.0~$\mu$s, 2.0~$\mu$s, and
3.0~$\mu$s.  The data used here were collected with the 
14 mV$\!$/fC gain setting and the 3.0~$\mu$s shaping-time setting.  The shaping-time setting was
maximised in order to reduce the impact of noise.  The gain setting is chosen
in order for the small expected signals to be visible.  The data were not compressed
on readout, and so the gain setting did not affect the data volume.  The ADC ASICs~\cite{domino-adc} digitise sixteen channels apiece at two million samples per second in
a continuous stream.  In what follows, the word ``tick'' denotes a
500~ns period of time corresponding to an ADC sample.  
With the gain and shaping-time settings as set, there are approximately 
$152\pm 18$ electrons per ADC count at the peak of a narrow pulse.   A voltage
offset of 200~mV is added to the output of the preamplifier to move the
baseline away from 0~mV for all channels, corresponding roughly to
600~ADC counts.  This offset is necessary in order to provide for the
readout of the bipolar signals on the induction-plane wires, as well
as to allow for signal recovery in case of noise or a downward
oscillation in the pedestal value.  The preamplifiers are DC coupled,
in contrast to the ProtoDUNE-SP preamplifiers, which are AC coupled.
The front-end ASICs and the ADC
ASICs are mounted on front-end motherboards (FEMB).  Eight of each
kind are mounted on each FEMB, for a total of 128 channels.

The digitised signals were sent to Reconfigurable Computing Elements
(RCEs)~\cite{rce} which triggered, buffered and formatted the data for
analysis and storage.  The RCEs transferred their data via Ethernet
to commodity computers running {\it artdaq}~\cite{artdaq}, a flexible
data-acquisition framework which provides hardware interfaces, event
building, logging, and online monitoring functionality.

Each triggered readout of the detector is 15000 ticks long and
starts between 4000 and 5000 ticks before each trigger, in order
to capture fully cosmic rays that overlay the
triggered interaction.  
The necessary buffering of the data is provided by the RCEs.
Because the disk-writing speed was limited to approximately 60~MB/s, the
detector readout was triggered at approximately 1~Hz.
Electronic noise in the detector and the small signals preclude the
use of zero suppression, and thus all ADC samples are recorded for all
triggered readouts.  Data are written in ROOT~\cite{root} format to a
single output stream by an {\it artdaq} aggregator process.    
The large electronics noise reduced the
maximum possible effectiveness of compression to a factor of $\approx 2$, with a large CPU penalty.
Therefore, no compression is applied, in order for CPU not to be a bottleneck in the
output data stream.

%% file: runningconditions_section.tex
\section{Running conditions}
\label{sec:run}

The nominal drift field in the DUNE FD design is
500~V$\!$/cm.  The data collected by the 35-ton Phase~II prototype were
taken at a field of 250~V$\!$/cm, however.  Compared with the nominal
field strength, the reduced field has several consequences.  
The drift velocity is reduced from a nominal~\footnote{The actual drift velocity differs from the nominal due to space-charge-induced local variations in the electric field.} 1.55~mm/\!$\mu$s to 1.04~mm/\!$\mu$s.  This lower drift velocity magnifies
the effects of the electron lifetime and diffusion on the collected
charge as a function of drift distance.  The lower field also increases the amount of
charge that recombines with the argon ions in order to make
scintillation light while decreasing the signals on the TPC wires.
The effects of space charge buildup due to slowly-moving positive ions
drifting towards the cathode are also increased by the lower drift
field.

\begin{figure}
\centering \includegraphics[width=0.8\textwidth]{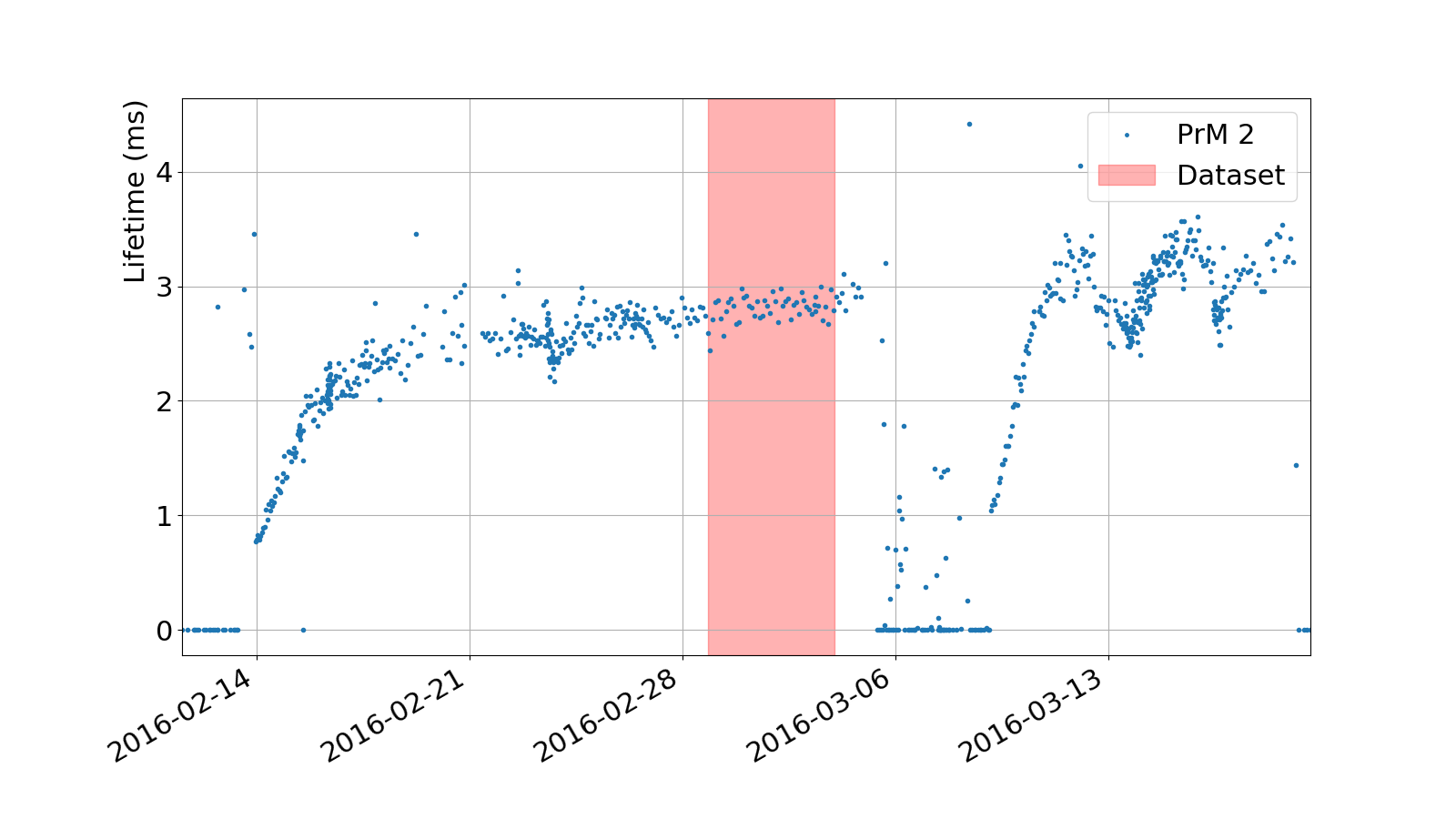}
\caption{The electron lifetime as measured by Purity Monitor~\#2, for the entire
  35-ton Phase~II run.  The shaded region corresponds the time region used in the offline electron lifetime analysis, which is described in section~\ref{sec:lifetime}.}
\label{fig:PrM2}
\end{figure}

The electron lifetime measured by the purity monitors was stable at
around 3~ms for the duration of the data-taking period.   The lifetime measured by Purity Monitor~\#2 is shown as a function of time in figure~\ref{fig:PrM2}.   The four purity monitors recorded different electron lifetimes.  The measured lifetime
decreased monotonically with height, with the top purity monitor measuring a lifetime $\sim$2~ms shorter than the bottom.
This stratification of the electron lifetime is attributed to the fact that
relatively pure, colder, filtered liquid argon is pumped into the cryostat near the bottom, and the recirculation
system's suction pipe is also located near the bottom.  In this arrangement, the liquid argon returned for recirculation
was colder than the average temperature in the cryostat, suppressing convective mixing and resulting in temperature
and purity stratification. The gas ullage above the liquid is predicted 
to have a much higher concentration of impurities than the liquid due to its higher temperature.  A detailed
computational-fluid-dynamic simulation of the flow, the temperature, and the estimated distribution of impurities
is given in~\cite{sdsu-cfd}.

Several short-lived operational issues, such as power outages and an exhausted
supply of liquid nitrogen, caused the electron lifetime to drop
temporarily.  The liquid argon purification system recovered the purity 
on the timescale of two days.   The main data-taking run was ended on March 19, 2016, when a metal tube
carrying gaseous argon to a recirculation pump broke due to metal fatigue
brought about by the vibration of the pump.  Air was pumped into the gas return line and mixed in with the liquid argon, resulting in a rapid loss of electron lifetime.  Noise and diagnostic data were collected after the incident but further purification of the argon was not attempted as sufficient data had been collected already.
Data used in the analyses presented here are selected
from only the high-electron-lifetime running periods.

The electronic noise was higher than anticipated in the 35-ton data.
In the worst case, a very high amplitude oscillatory noise with an amplitude of 200~ADC counts (30,400 electrons) per channel was seen throughout the detector, and
corresponded to a self-sustaining ``high-noise" state.  The detector
entered this state spontaneously, though only when the drift field was
turned on and the anode wire planes were biased.  The high-noise state
could be cleared by removing power from the front-end boards,
restoring power to them, and re-initializing them.  It was found in
the course of the run that switching off the front-end boards of 
APA~1 helped to prevent spontaneous triggers of the high-noise
state.  Section~\ref{sec:rawdata} describes the characteristics of the
data when not in the high-noise state.

A number of wires were not read out for part or all of the run, due to
both wire breakage and issues with the readout electronics.
Twenty-seven wires were broken during APA fabrication and testing, all
of which are induction-plane wires.  Of these, ten remained
mechanically secure but their electrical connections to the front-end
electronics were severed during a thermal test.  The long wire
segments on the sides of the breaks away from the electronics were
jumpered to their neighbors in order to preserve the electrostatic
configuration of the APAs.  The remaining 17 broken wires were
removed.  

During initial commissioning following installation and before the
first cooldown, 74 electronics channels out of a total of 2048 were identified as malfunctioning using calibration pulser signals.
Seven front-end ASICs stopped working after the first cold power
cycle, comprising 112 channels, which were not read out for the duration
of the run.  Eight ADC ASICs, comprising 128 channels, could
not be synchronised correctly and thus they also did not contribute data for the
duration of the run.  Two front-end motherboards, comprising 256
channels, lost their low-voltage power due to a short circuit on the flange board partway through the run.  The front-end motherboards serving
the shortest APA, with 512 channels, were turned off in order to reduce
the frequency of transitions into the high-noise state.
A total of 28\% of the TPC channels were  % number from run 17089
not functioning or not being read out at the end of the run.  
Nonetheless, enough data were
collected in order to test the design choices and meet some of the goals of the
prototype.

%% file: rawdata_section.tex
\section{Raw data characteristics}
\label{sec:rawdata}

\begin{figure}[ht]
\includegraphics[width=\textwidth]{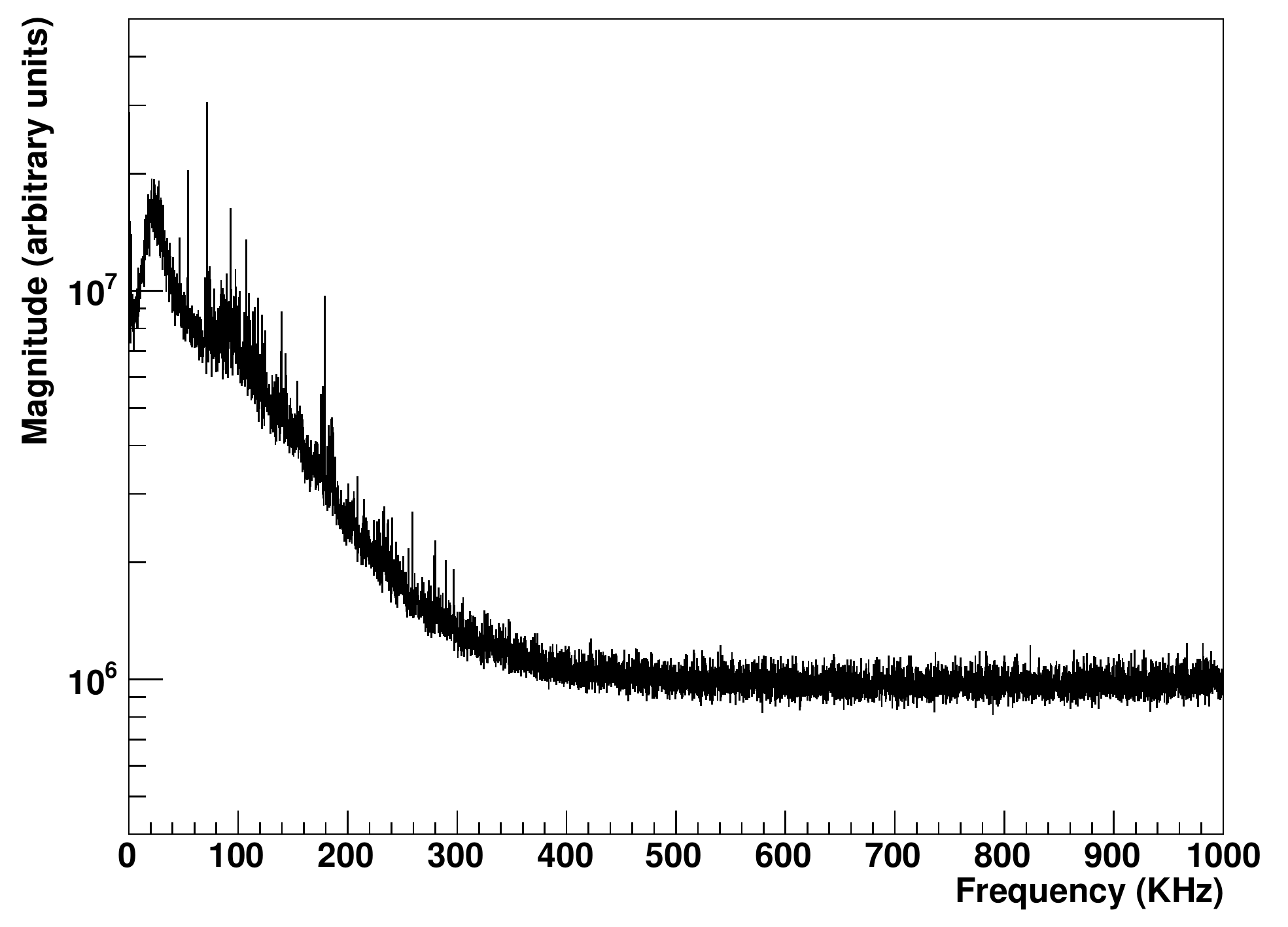}
\caption{\label{fig:noise_spectrum_1d} The average magnitude of the Fourier transform of the ADC 
values read out for all good channels in the 35-ton prototype, for a single 15000-tick event in a  low-noise run. The spectrum shows numerous noise peaks
superimposed above a white noise background, which is attenuated at high frequency due to the shaping time of the preamplifier.}
\end{figure}

When not in the high-noise state, the standard deviation (RMS) of the digitised signal values was in the range of 20-30 ADC counts (3040-4560 electrons). 
A frequency spectrum of this
noise is shown in figure~\ref{fig:noise_spectrum_1d}.  The noise consists
of correlated and uncorrelated components, both of which are functions of time.
An analysis of the correlations of the ADC values determined that
correlations were strong within the 128 neighboring channels that
share a FEMB.  This particular component of the noise is 
ascribed to a voltage regulator on the FEMB.
The correlated characteristic of the noise is used
in the coherent noise subtraction step, described in
section~\ref{sec:dataproc}.  Additional sources of noise were identified to have arisen
from a feedback loop between the low-voltage supply regulators and the time delays in the
long cable runs from the power supply to the detector, as well as incomplete grounding
isolation due to the conductivity of the steel-reinforced concrete structure supporting the cryostat.

The data are also affected by bit-level corruption.  In a fraction of
ADC samples, which depends on the temperature, the channel and the
input current, the least-significant six bits (LSB6) of the ADC could be erroneously reported
as {\tt 0x0} or {\tt 0x3F}.  These values are referred to as ``sticky codes''.
If LSB6 is erroneously {\tt 0x0}, 
then the number represented by the upper six bits is one greater
than it would be if LSB6 had not been in error. When the LSB6
is erroneously {\tt 0x3F}, then the number represented by the upper
six bits is one less than if LSB6 had not been in error.  The probability
that LSB6 will be in error  depends strongly on the proximity of
the true input value to the boundary in which the result would
be {\tt 0x0}.  These fractions of ADC
samples varied from 20\% to 80\%, depending on the factors mentioned
above.   Some ADC samples for which LSB6 is
{\tt 0x0} or {\tt 0x3F} are in fact correctly digitised.  The values {\tt 0x00}
and {\tt 0x3F} for LSB6 are the most common sticky codes but others have been observed,
such as {\tt 0x01}.
Procedures for flagging and mitigating this corruption are described in 
section \ref{sec:dataproc}.

\begin{figure}
  \centering
  \includegraphics[width=7.5cm]{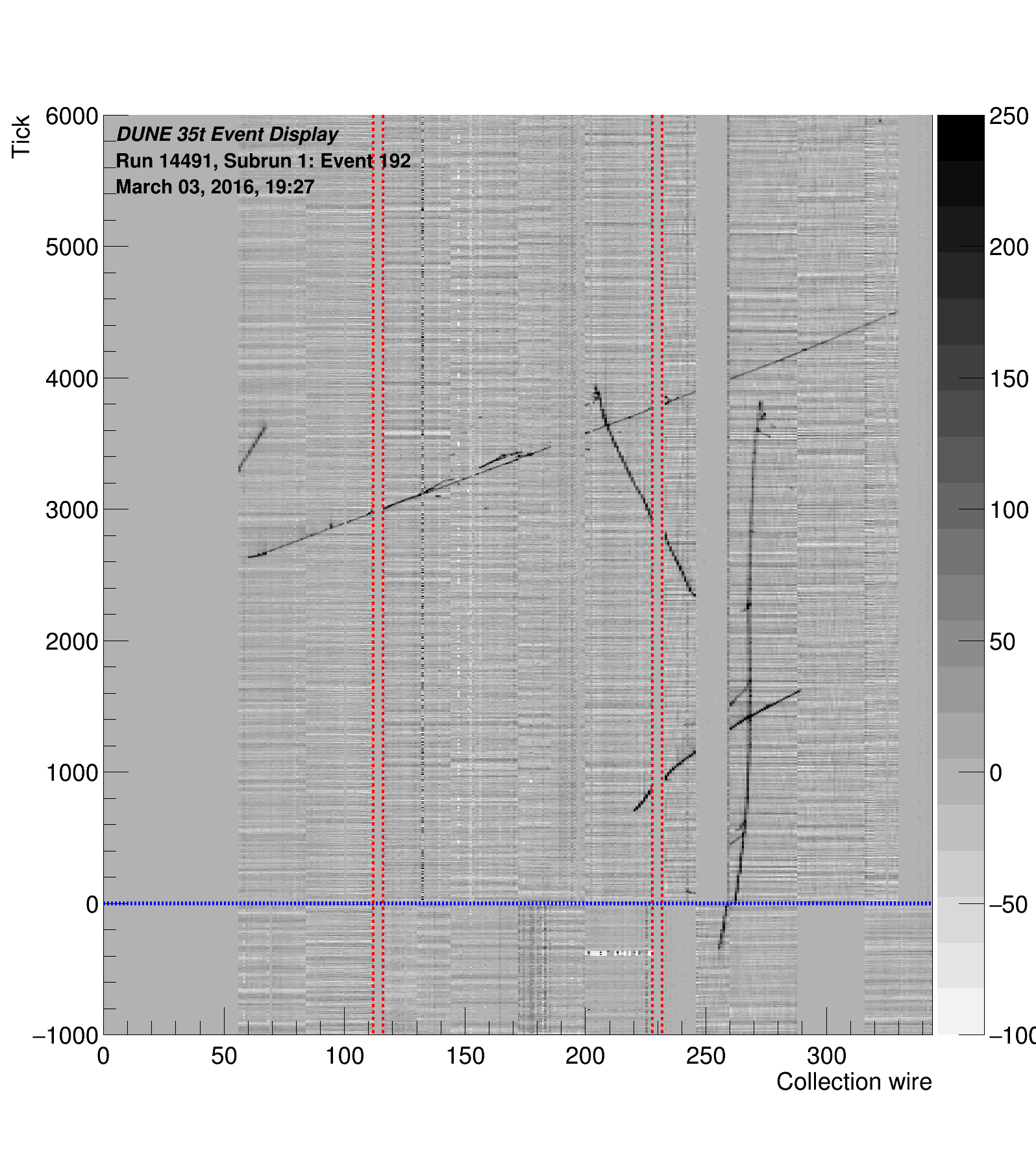}
  \includegraphics[width=7.5cm]{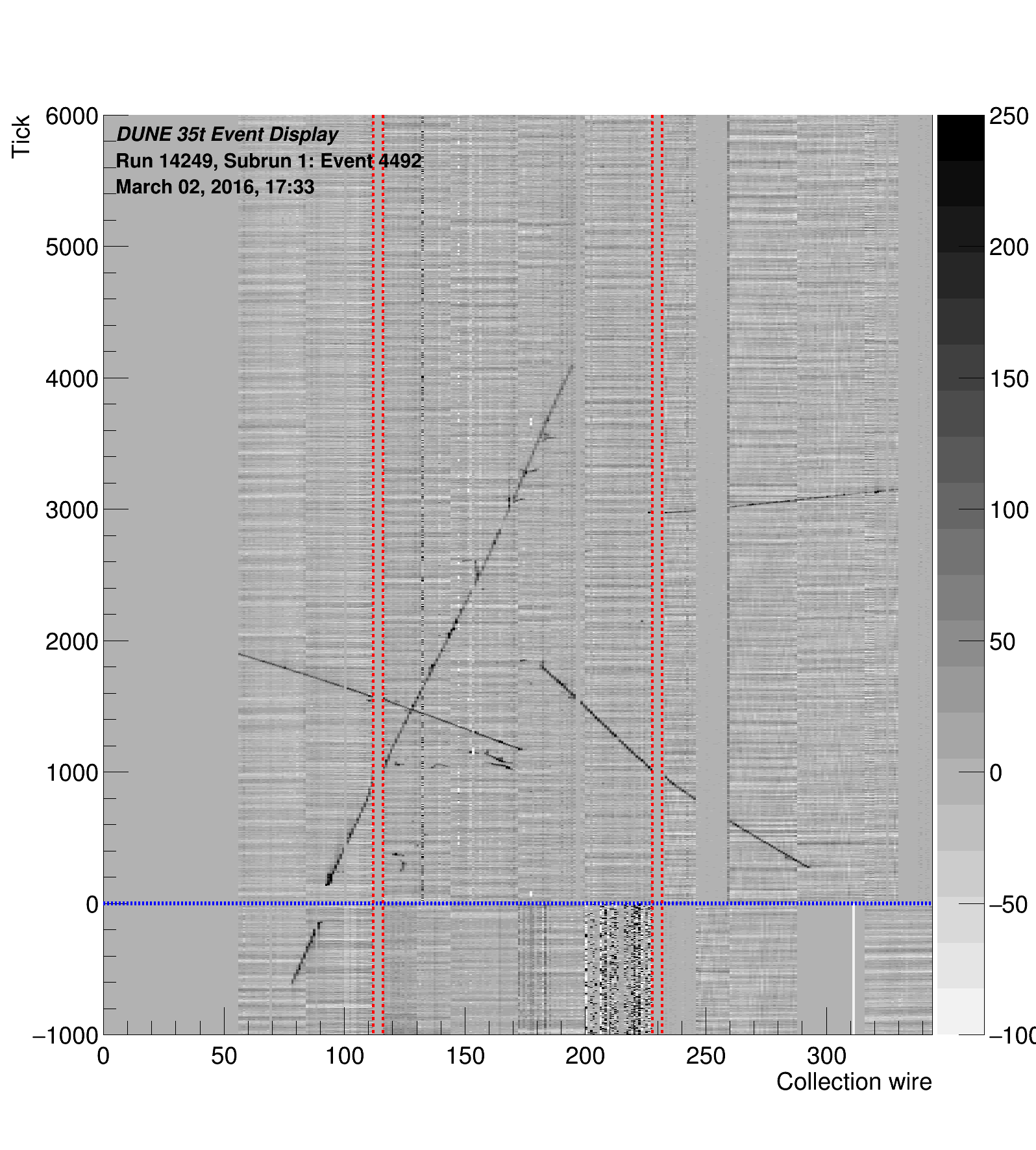}
  \caption{The two displays of triggered readouts show collection-plane raw data.  
  Multiple cosmic-ray tracks are visible in both events. Darker pixels indicate higher 
  ionization deposits. The horizontal axes indicate the wire number and the vertical axes indicate the time at which the signal on the wire is digitised. A study of the tracks
    which pass across gaps between the APAs (indicated by dotted red lines) is the
    subject of section~\ref{sec:zgap}.  Offsets are visible as tracks cross through the APAs (indicated by dotted blue lines). Correcting for $t_0$ yields connected tracks, as discussed in section~\ref{sec:apagap}.}
  \label{fig:evd_crossing}
\end{figure}

Figure~\ref{fig:evd_crossing} shows two examples of the raw data for triggered events. Multiple 
cosmic-ray tracks are visible in both events. 

%% file: dataprocessing_section.tex
\section{Data processing}
\label{sec:dataproc}
%noise reduction, stuck code removal, deconvolution, hit-finding, and performance of this (s/n for layers)

The first stage in processing is data preparation: the raw data are
unpacked, pedestals are subtracted, noise and other issues are
mitigated and deconvolution is performed. The steps in this
preparation are detailed below. The initial processing is performed
independently for each readout channel. Channels flagged as bad are
not processed.

The first step in the data preparation process is
data extraction. The raw data for each channel are
unpacked and converted to floating-point format and the most-recent
pedestal (evaluated in dedicated runs and stored in a database) is subtracted. The extracted
data include an ADC value for each of the 15000 readout ticks for each
channel in each event. In addition, a flag is set for each tick to
inform downstream algorithms of possible issues in the
measurement. The flag is either cleared or set to one of four warning
values based on the 12-bit ADC value. The possible values for the flag are underflow
(ADC value={\tt 0x0}), overflow (ADC value={\tt 0xFFF}), low sticky code (LSB6={\tt 0x0}) or
high sticky code (LSB6={\tt 0x3F}).  Other sticky codes are not flagged.

\begin{figure}[ht]
\begin{center}
\includegraphics[width=0.48\textwidth]{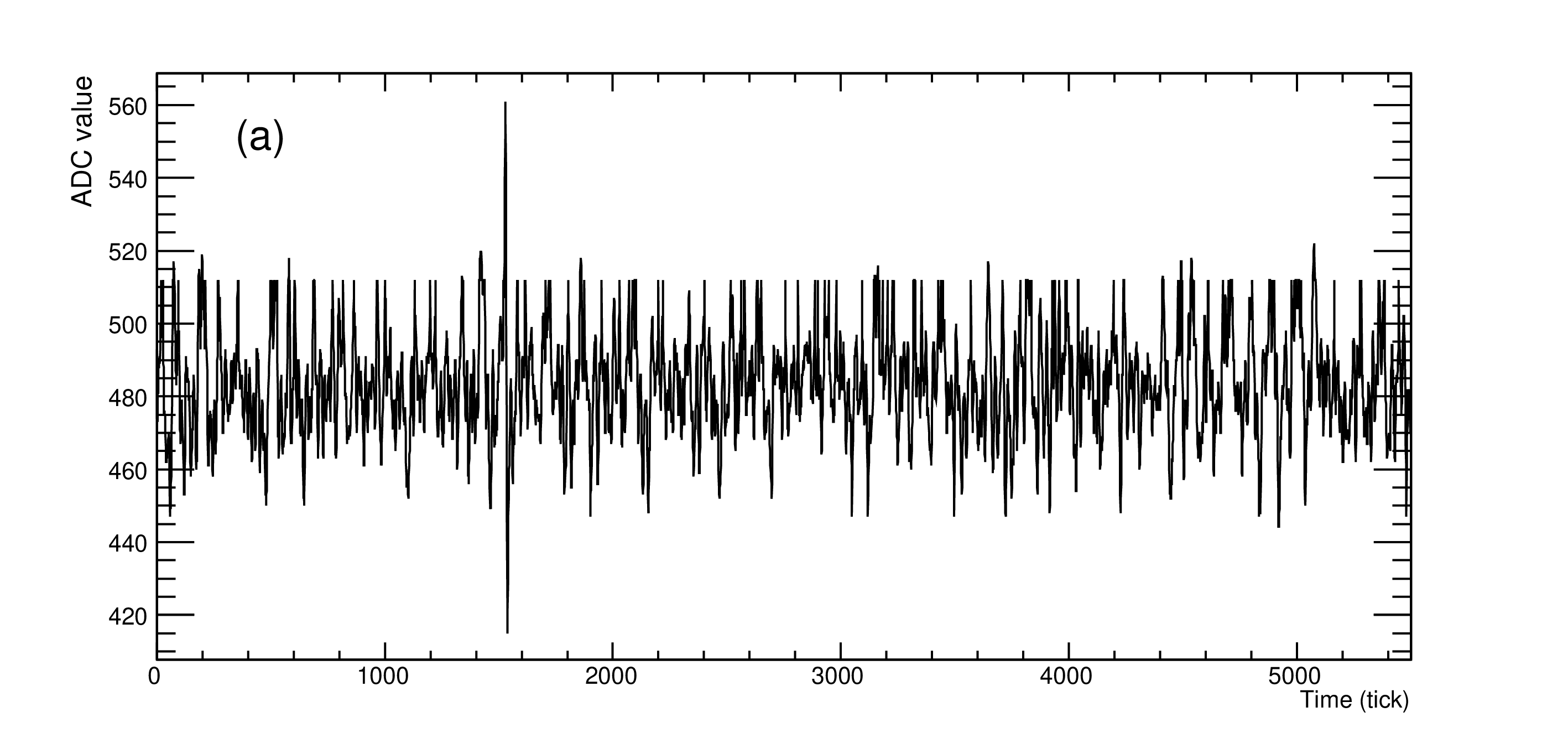}
\includegraphics[width=0.48\textwidth]{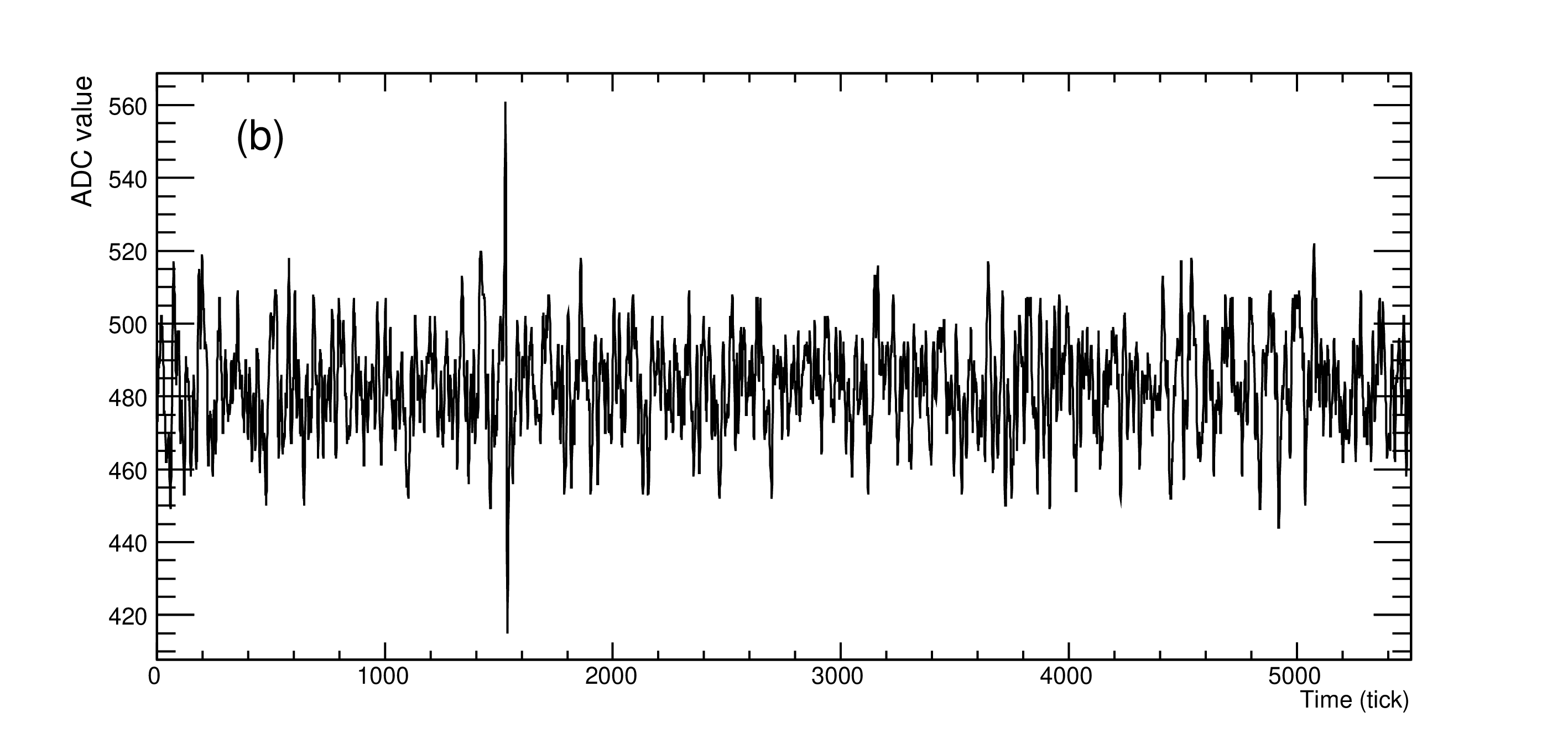}
\includegraphics[width=0.48\textwidth]{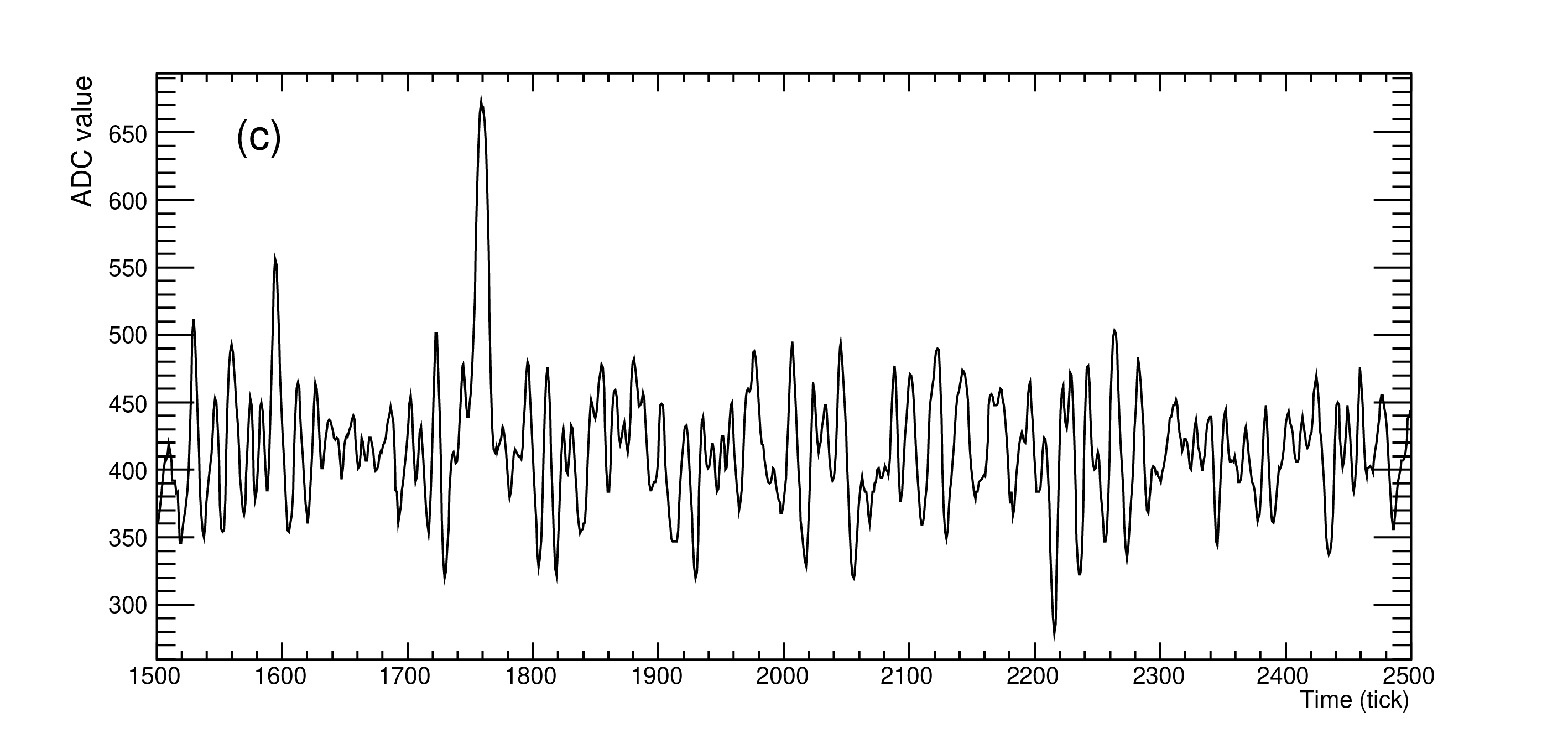}
\includegraphics[width=0.48\textwidth]{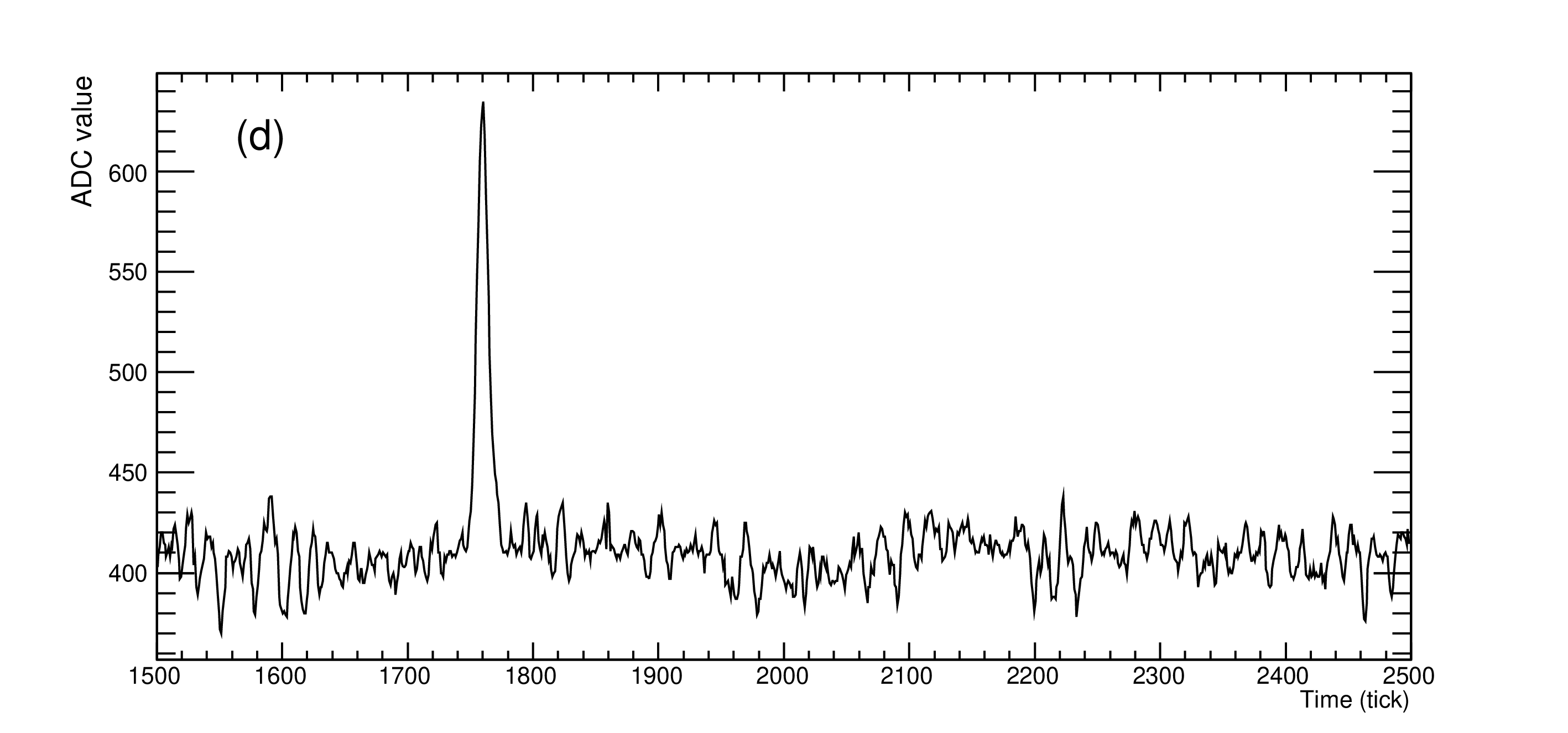}
\end{center}
\caption{\label{fig:dataproc} Example waveforms at differing stages of data processing.  Waveform (a) shows a raw waveform from an induction-plane channel with sticky codes present.  Waveform (b) is the same as (a), but with the sticky codes mitigated.  Waveform (c) is a different waveform from a collection-plane channel, before correlated noise removal.  Waveform (d) is the same as (c), but after correlated noise removal.  All waveforms have visible signals present in addition to the noise.}
\end{figure}

The next step is ADC mitigation, which attempts to correct for the bias
and poor resolution that would follow from direct use of ticks with
sticky ADC codes. The extracted values for ticks with sticky code flags are discarded
and replaced with the values obtained by linearly interpolating
between the values from the nearest preceding and following ticks
that do not have sticky codes. If there are no ticks without sticky codes
on one side, then the value on the other side is used. No replacement is made if the tick is
in a series of more than five ADC values with sticky codes. Where a replacement is
made, the ADC flag is set to a new value to indicate that an
interpolation (or extrapolation) has been performed.  Figures~\ref{fig:dataproc} (a) and~\ref{fig:dataproc} (b) show 
a waveform with sticky codes before and after mitigation.

Correlated noise removal is the next step in the processing. As
discussed above, a strong correlation is observed between the noise in
the 128 channels that are processed by each FEMB. Separately for 
each tick and readout plane, a median ADC value is evaluated for all
contributing channels. 
 The noise is estimated with the median
rather than the average to reduce the influence of the signal on the noise, and also
to reduce the impact of other tails such as that from pedestal mismeasurement.
The median is then subtracted from the ADC value
for that tick in each channel within the corresponding group, indexed by FEMB and plane.
ADC values corresponding to sticky codes
are corrected before being included in the calculations of the median values. Figures~\ref{fig:dataproc} (c) and~\ref{fig:dataproc} (d) show 
a waveform before and after correlated noise subtraction.

The channels read out by an FEMB
within a plane typically correspond to adjacent wires,
and so ticks with signals from charged particles are likely
to contribute to the noise estimate. 
An improved version of this algorithm, which suppresses the contribution of signals
to the background estimate~\cite{Acciarri:2017sde}, was developed by the MicroBooNE
Collaboration.  The relatively poor S/N ratio in the 35-ton prototype
however makes such a modification less effective.

The next step in signal processing is frequency-domain filtering and deconvolution,
which are combined in one step.  For each channel, an FFT is
performed on the raw ADC values as a function of time to obtain a
frequency-domain representation of the data, which is then multiplied
by the product of the deconvolution kernel and a noise filter.  The
deconvolution kernel is defined separately for induction-plane signals
and collection-plane signals, and is the reciprocal of the FFT of the
simulated response of the detector and electronics to a single impulse
of charge arriving in a very short time~\cite{garfield}.  Poles in the kernel are set
to a maximum value so as not to emphasise noise that coincides with a
zero in the detector response. The noise filters, one for
induction-plane channels and one for collection-plane channels, were
constructed from representative waveforms containing visually
identifiable signals from tracks traveling roughly perpendicular to
the drift field.  Portions of the waveforms corresponding to
identifiable hits were removed and the spectrum of the remaining
waveform was calculated to estimate the noise-only spectrum, and the
regions in time near the hits were used to calculate the spectrum of
the signal.  The noise filter is then a smoothed representation of $s/(s+n)$ as a
function of frequency.  Generally, frequencies between 20 and 120 kHz
are retained while others are filtered out.

The final step in data preparation is identification of regions
of interest (ROIs), i.e., consecutive ticks in each channel that appear to
hold signals from charged particles. Only these regions are retained
for downstream processing. An expected noise level is assigned for
each plane orientation and an ROI is constructed where the
deconvoluted signal for a channel exceeds three times the noise level
and extends in either direction until the value for a tick falls below
the noise level. The ROI is then extended by 50 ticks on each end.

The typical peak signal size for tracks traveling parallel to the wire plane is 100 ADC counts (15200 electrons) on the collection plane and 45 ADC counts (6840 electrons) on each of the two induction planes.  The noise levels are characterized by the standard deviation of the waveform values sampled on each tick and are between 20 to 30 ADC counts (3040 to 4560 electrons) per tick.
The peak S/N ratios are therefore near~5 for collection planes and around~2 for the induction planes. 
This means that the the hit-finding and 2D track-finding efficiencies are higher in the collection plane than in the induction planes. Nonetheless, analyses presented below rely on 3D reconstruction, which is possible sufficiently often to complete the measurements.

%% file: hitfinding_section.tex
\section{Hit finding and track finding}
\label{sec:hitstracks}

Three hit-finding algorithms, called the Raw Hit Finder (RHF), the
Gauss Hit Finder (GHF)~\cite{Acciarri:2017hat},  and the Robust Hit Finder (BHF), are in use in the analyses presented here.  
The RHF and GHF are standard algorithms used in other LArTPCs, whereas the BHF was developed
specifically for the conditions of the 35-ton prototype.
All three hit finders and the tracking algorithms used here make use of the LArSoft toolkit~\cite{larsoft}.

The RHF operates on pedestal-subtracted
but otherwise un-deconvoluted or filtered raw ADC values and applies
thresholds to identify the times and charges of the hits.

The GHF uses deconvoluted and filtered data and proceeds in two steps.
The first step is a peak-finding algorithm which applies a threshold
to find a peak, and seeks troughs between neighboring peaks to count
the number $n_{\rm{gauss}}$ of nearby peaks within a region of
interest, which itself is determined by thresholds and a minimum
number of ticks in the region.  A function, constructed from the sum
of $n_{\rm{gauss}}$ Gaussian functions, is then fit to the
deconvoluted data in the region of interest.  The reconstructed hit
information consists of the Gaussian fit parameters and also the sums
of the deconvoluted-filtered ADC values corresponding to the time
windows for the hits.  The time of the reconstructed hit is defined to
be the time at which the gaussian fit has its maximum.  

The BHF is an algorithm that does both hit reconstruction and 2D track
reconstruction using collection-plane raw digits and muon counter
information~\cite{Thiesse:2017qom}.  Hits are sought in two-dimensional
``roads'' defined by the region in the detector consistent with a track
passing through CRCs with time-coincident hits.  Stuck codes are mitigated and noise is filtered in the time
domain.  Hits are identified by the significance of the excursion of the waveform
from the pedestal in units of the standard deviation of the waveform outside the
candidate signal region.

Tracks are found using three methods: The Counter-Shadow Method (CSM),
the Projection Matching Algorithm (PMA)~\cite{pma}, and the Track Hit
Backtracker (THB).  The CSM algorithm seeks hits within the areas
geometrically bounded by the CRCs in space and time,
assuming that the track is a straight line.  Wires with multiple hits
within the counter shadow are not used, as they may be noisy or have
hits from delta rays.  The mean square residual per hit from a line
fit is required to be less than 1.0~cm.

The PMA method starts with clusters of hits in each of the three views
made by the TrajCluster algorithm~\cite{trajcluster}.  The principle
of TrajCluster is similar to that of a Kalman Filter~\cite{kalman} to
identify particle trajectories that may include scattering in the
dense liquid argon medium.  Hits are added to clusters based on their
consistency with the trajectory established by previous hits on the
cluster, based on the local direction of the trajectory and the
expected variation in position and angle to the next hit.  This method
effectively rejects delta rays and identifies kinks in tracks.  The
PMA algorithm then identifies matching parts of the 2D clusters, and
fits the projections of 3D track hypotheses to the data in the three
2D views.  The PMA algorithm can successfully reconstruct tracks with
data from two planes, up to the ambiguities introduced by the wrapped
wires that are resolved by the third plane's data.

The THB algorithm was developed to purify the hits found by
the BHF and to recover charge signals that were missed because
they were below the hit-finding threshold.   If a sequence of BHF 
hits left by a throughgoing cosmic-ray track is missing one or more 
expected hits with found hits on either side, then the charge 
deposited by the track is nonetheless 
assumed to be present for the channels missing hits.  
The locations are interpolated from
neighbouring hits, and the charges are computed from the waveforms as
if the hits had been found. More details are available in
Ref~\cite{Thiesse:2017qom}.

\begin{figure}[ht]
\centering \includegraphics[width=\textwidth]{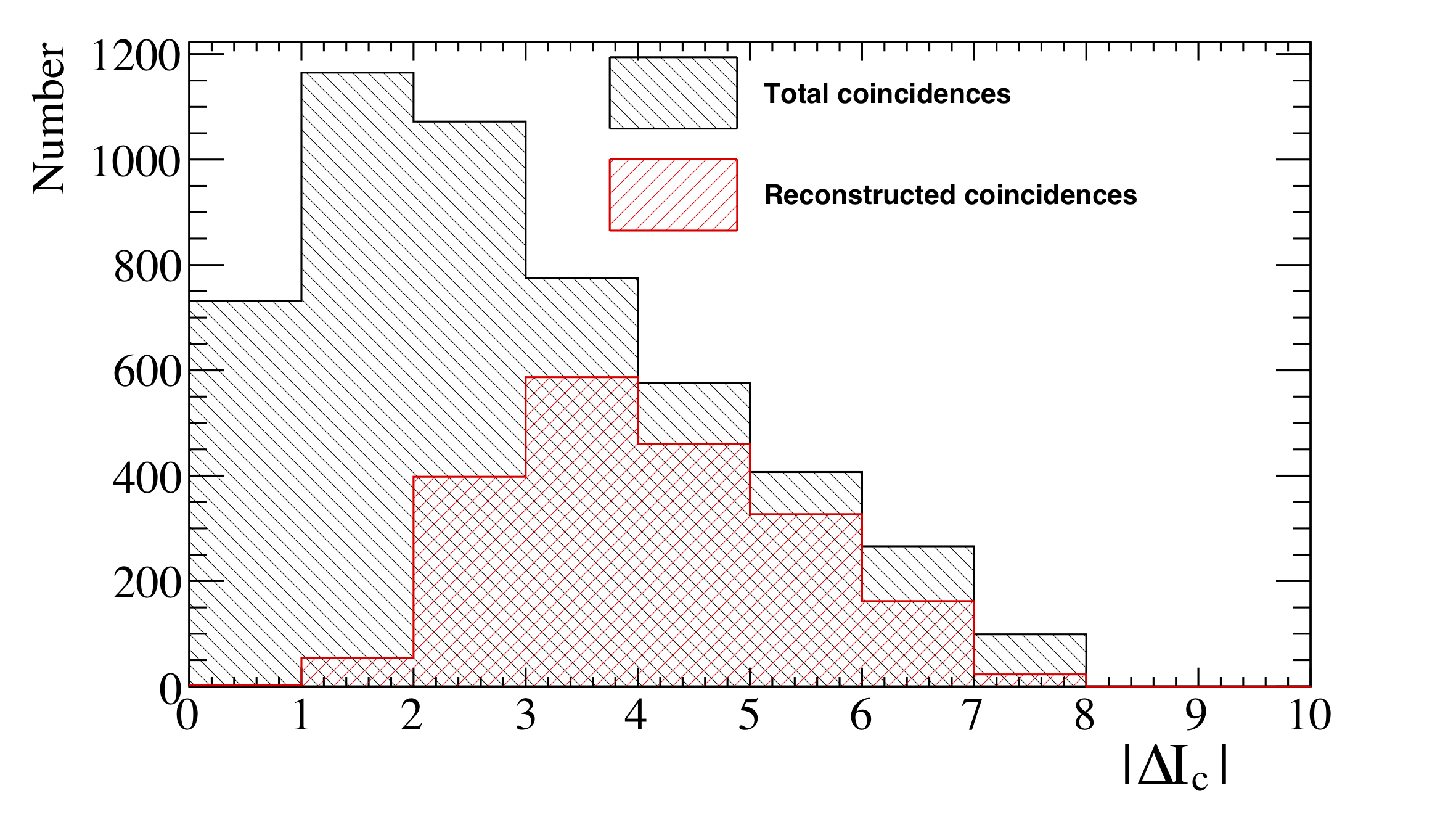}
\caption{A comparison of the number of measured counter coincidences,
  and the number of associated tracks which are reconstructed with the PMA algorithm, as a function of the absolute value of the difference in counter $x$-position indexes.  }
\label{fig:CounterRecoEffic}
\end{figure}

The combined track reconstruction and selection efficiency is estimated by comparing the number of counter coincidences against the number of
reconstructed tracks that meet selection criteria.  In order to be selected, a track
must contain at least 100 collection plane hits out of an expected 300.  
Figure~\ref{fig:CounterRecoEffic} provides an estimate of the tracking efficiency as a function of track angle.  The distribution of East-West counter-pair coincidences is shown, along with those that have matching tracks reconstructed with the PMA algorithm, as functions of the absolute value of the difference in the counter indices $|\Delta I_c|$.  The counter
indices increase along the $x$ direction.   Tracks that pass through counters with the same index travel in planes nearly parallel to the APA plane.  The
tracking efficiency is low, especially for tracks that pass through opposing counters that have small
differences in their $x$ locations.  The reason for this inefficiency is that drifting electrons from
ionization along these tracks arrive at the anode at similar times and the correlated noise removal algorithm
suppresses both the noise and the signal.

%% file: relativealignment_section.tex
\section{Relative alignment of the CRCs and the TPC}
\label{sec:counteralignment}
The CRCs are used to search the TPC data for signals corresponding to cosmic
rays that pass through pairs of counters, as well as to determine the event
time $t_0$.  Coincidence triggers were formed in order to select cosmic rays traveling
from east to west or vice versa, and from north to south or vice versa, using pairs of CRCs
on facing sides of the cryostat.  In order for the fiducial cuts to correctly isolate the
relevant signal region from the background region, the relative
positions of the TPC and the CRCs must be determined.  Because the TPC is sealed in the
cryostat and is thus inaccessible, this alignment is accomplished with cosmic rays.
Unlike the other analyses described in this paper, only collection-plane data are used
for the alignment studies.

\subsection{East-West CRC alignment}
\label{sec:ewcounters}
\begin{figure}[ht]
\centering
\includegraphics[width=0.7\textwidth]{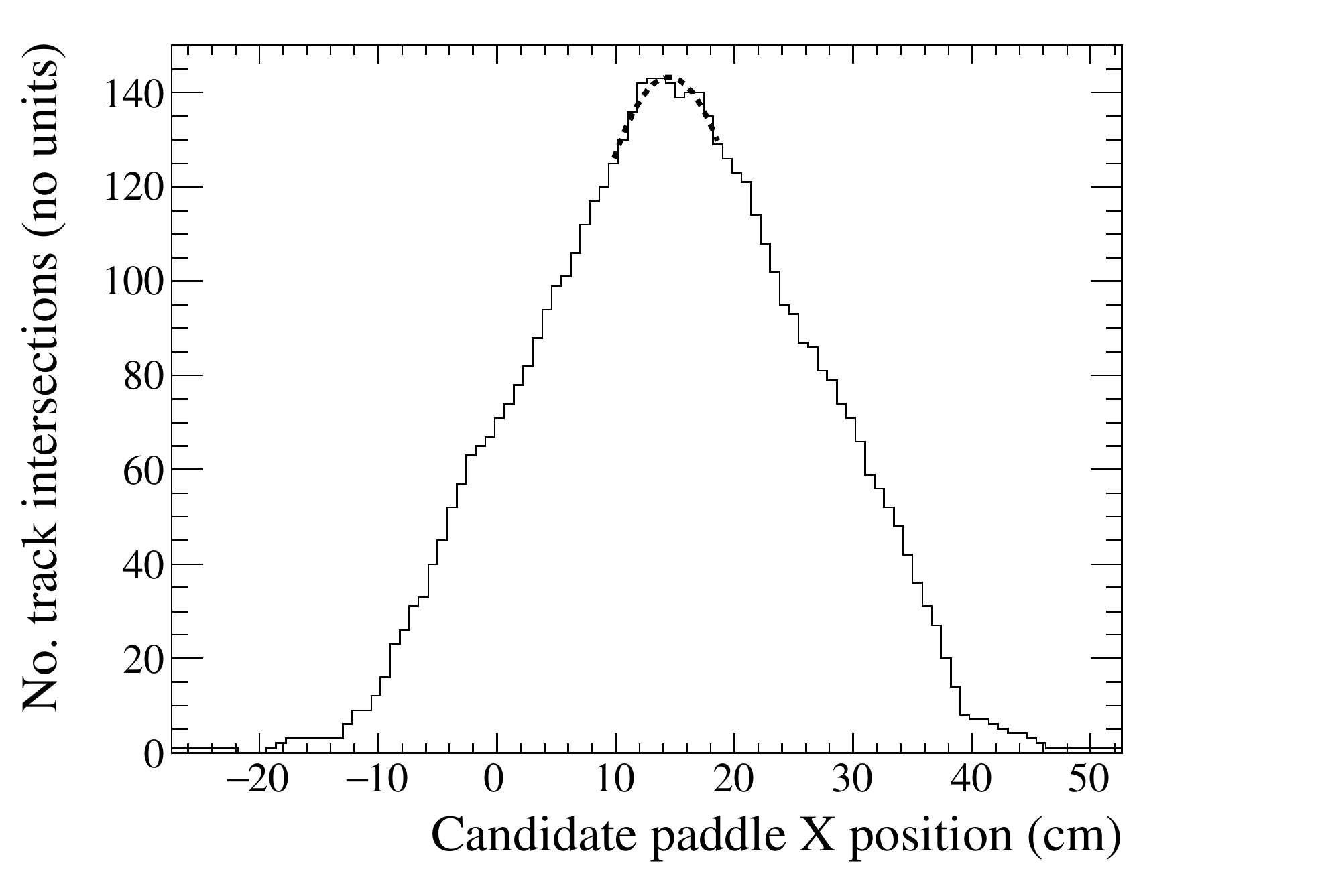}
\caption{The result of the  search which counts the number of
  reconstructed track intersections for a single East-West muon counter pair. % number 30
  In order to smooth out statistical fluctuations near the peak, a locally parabolic function
  is fit in the neighborhood of the peak and the location of the function's peak
  is used as the measured value of the counter position.
}
\label{fig:FitPositionEWCounter30}
\end{figure}

Data triggered by East-West CRC pairs is the most widely used in our 
analyses.
Therefore,  once
TPC data became available, estimates of the East and West counter
positions were made using collection-plane measurements of cosmic-ray
tracks that triggered directly opposite CRC paddles.  
Tracks triggered by East-West counter pairs are traveling roughly parallel 
to the anode planes.
The tracks in the
TPC are selected using the CSM, and CRC paddle positions are measured
by maximising the numbers of tracks which extrapolate to intersect
the paddles as functions of the assumed paddle positions.  Only the
$x$ coordinate of paddle pairs is thus measured -- the $z$ positions 
are taken from measurements of the outside of the cryostat support walls,
and since only collection-plane wire data are used, $y$ is not measured.
The expected distribution of the number of tracks intersecting a counter
as a function of its location in $x$ is approximately triangular, convoluted with a Gaussian
which accounts for multiple scattering and detector resolution.  An example
distribution is shown in figure~\ref{fig:FitPositionEWCounter30}.  The
statistical precision on the extraction of the location of the
distribution's peak is improved
by fitting a smooth function in the neighborhood of the peak and using the
peak of that function.  The resulting fit locations are used in the other
analyses presented in this paper.

\subsection{North-South CRC alignment}
\label{sec:nscounters}

Tracks triggered by North-South counter pairs travel roughly in the 
$x$~direction (perpendicular to the anode planes).  
They can be used to measure the positions of the North-South counters relative to the TPC, 
with the strongest constraints in the $z$~direction.
The signals left by these tracks are weaker in the induction-plane channels
than for tracks passing at a larger angle with respect to the electric field
due to the cancellation of nearby positive and negative components of the bipolar signals,
reducing the signal-to-noise level for these tracks relative to that reported in section~\ref{sec:dataproc}.
For this reason, the analysis presented here to align the North-South counters uses only collection-plane
data, and so the $y$ coordinate of charge deposition in the TPC is not
measured.   Extrapolations of tracks triggered by North-South counter pairs are also affected 
by distortions due to space charge buildup in a way that is difficult to constrain with the data.
Therefore, a simpler, more robust method was devised in order to constrain just the $z$ locations
of the North-South counters.

\begin{figure}[ht]
\centering \includegraphics[width=\textwidth]{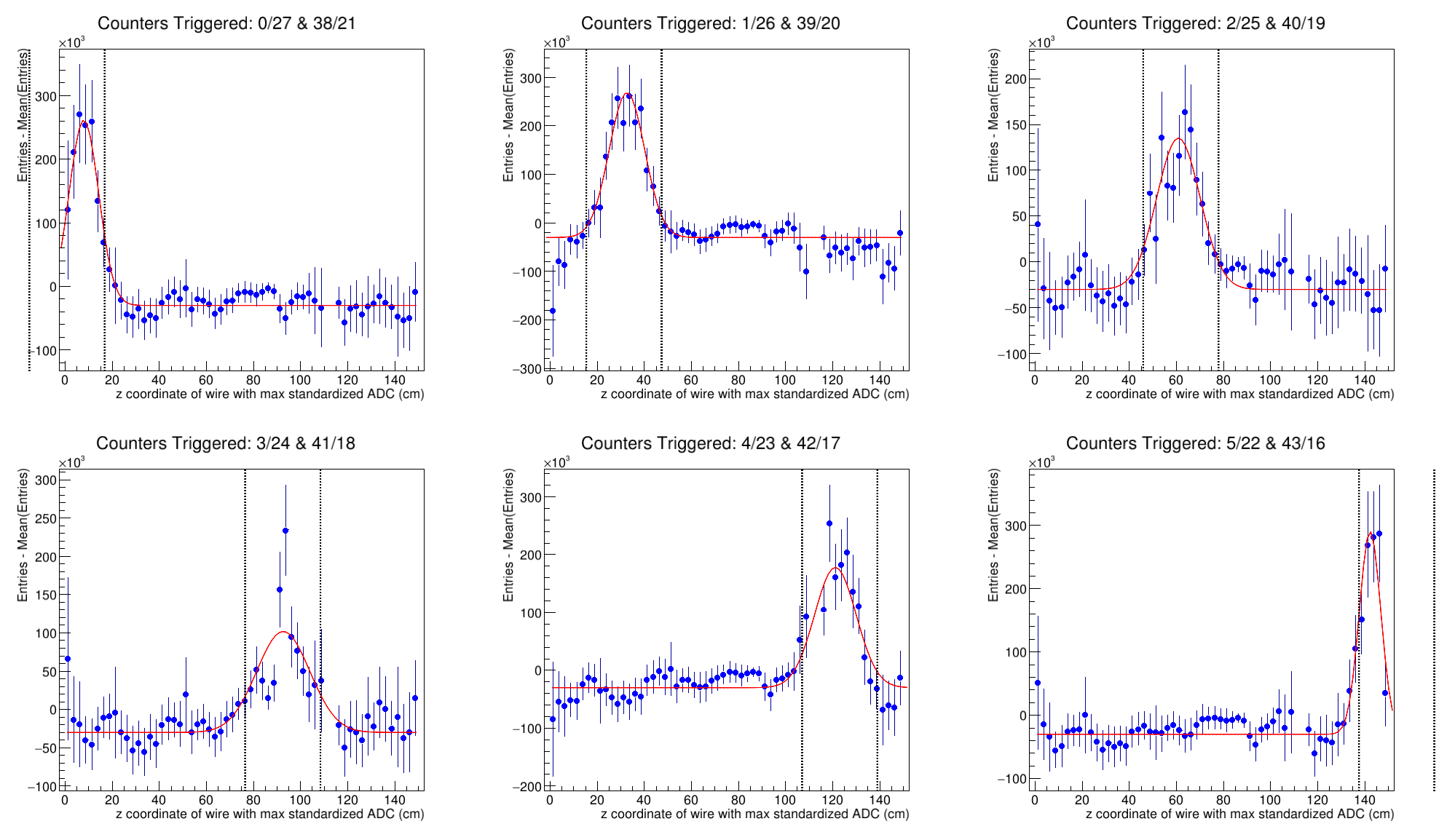}
\caption{Measured $z$ position of North-South muon counters by finding
  the wire with largest standardised ADC value for each TDC
  tick. Distributions are fit to Gaussians plus constant offsets to obtain central
  values.  The dotted lines indicate external survey measurements of the $z$ boundaries of
  the corresponding muon counter pairs.}
\label{fig:counterz}
\end{figure}

For every TDC tick on every collection plane wire, the $z$ coordinate of the
wire with the greatest standardised ADC, defined to be the ADC value's difference from 
the mean divided by the RMS of the ADC values on the wire, was
histogrammed. Due to random fluctuations in the baseline noise in the
absence of signal, each wire is equally likely to contain the maximum standardised
ADC value. But in the presence of signal, we expect the wire
containing the signal to be chosen more frequently in this selection
due to an excess of charge deposited, and, hence, we can determine the
$z$ location of the triggered muon counter pair. 
Figure~\ref{fig:counterz} shows the results
for triggers from pairs of counters directly opposite one another
(at the same nominal $z$ position), compared with external survey measurements of the counter locations.
A Gaussian plus a constant function describes the  observed distributions of the $z$ locations, and
for the four central counter pairs, is used to determine the best-fit $z$ locations.  The outermost
counter pairs extend beyond the TPC dimensions and  thus their distributions are truncated.

%% file: zgap_section.tex
\section{$Z$-gap crossing tracks}\label{sec:zgap}

One of the primary motivations for the design of the 35-ton TPC was to
test the performance of its modular anode plane assemblies.  In the
35-ton TPC, as in the FD design, multiple anode assemblies
are joined together to read out a shared volume of liquid argon.  Some
of the particles passing through the detector will traverse the
vertical gaps between the APAs (the $z$-gap), and some will traverse the horizontal
gap (the $y$-gap).  The subset of the 35-ton dataset consisting of muons which pass
across the face of APAs and which therefore deposit charge on
neighbouring APAs is discussed in this section.  Examples of such
tracks can be seen in the event display in
figure~\ref{fig:evd_crossing}.

The track segments from neighbouring TPCs can be used to measure the
gap between the corresponding APA frames.  This is performed by
minimising the total $\chi^2$ summed over all track segments as a function of APA gap hypotheses.
The offset from the assumed value can be determined for the vertical gaps
between the following pairs of TPC volumes: 1 and 3, 1 and 5, 3 and 7, and 5 and 7.  
The locations of these TPC volumes are shown in figure~\ref{fig:tpcphotonumbering}.
The number of particles depositing sufficient charge in the
short drift volume in the data sample was too low to make a statistically significant
measurement of the gaps between TPCs in this region.

The alignment of tracks crossing APA boundaries is sensitive to
offsets in both the $x$ and $z$ directions; tracks at multiple angles with respect to the
APA plane are required to fit for both of these offsets for each gap.  
The offsets measured by applying this method to each of the gaps are presented in 
Table~\ref{tab:APAGapOffsets}, along with the nominal distances between collection-plane wires in neighbouring TPC volumes.
The uncertainties shown in the table are statistical only; the effects of
systematic uncertainties are not considered and they are assumed to be
negligible in comparison.    The correlations in the uncertainties between
the $x$ and $z$ offsets in joint fits to both variables is very small.
Table~\ref{tab:APAGapOffsets} also lists the sums of the gap offset measurements for
the gap between TPCs~1 and~3 added to the gap offset measurements for the gap between TPCs~3
and~7, compared with similar sums with TPC~5 as the intermediate path.  Comparing these
sums provides both a measurement of the consistency of the method and an estimate of
the constancy of the gap width offsets as functions of $y$.

\begin{table}
  \caption{The measured offsets with respect to the assumed gap width between the APAs, in $x$ and $z$, along with the number of tracks utilised in each sample.}
  \label{tab:APAGapOffsets}
  \centering
    \begin{tabular}{l  c  c c  c  c  c }
      \toprule
          &           & Assumed    &             & \\
      Gap & Direction & width (cm) & Offset (cm) & \# Tracks \\         
      \midrule
      TPC 1/TPC 3 & $x$ & 0 & $-0.377 \pm 0.006$ & 335  \\
      TPC 1/TPC 5 & $x$ & 0 & $-0.252 \pm 0.002$ & 1810 \\
      TPC 3/TPC 7 & $x$ & 0 & $-0.16\ \, \pm 0.01\ \,$   & 88   \\
      TPC 5/TPC 7 & $x$ & 0 & $-0.286 \pm 0.002$ & 2612 \\
      \midrule
      TPC 1/(3)/TPC 7 & $x$ & 0 & $-0.537 \pm 0.010$ \\
      TPC 1/(5)/TPC 7 & $x$ & 0 & $-0.538 \pm 0.003$ \\
      \midrule
      TPC 1/TPC 3 & $z$ & 2.08  & $-0.18\ \,\pm 0.02\ \,$  & 335  \\
      TPC 1/TPC 5 & $z$ & 2.08  & $\ \ \,0.131 \pm 0.007$ & 1810 \\
      TPC 3/TPC 7 & $z$ & 2.08  & $\ \ 0.10\ \  \pm 0.03\ \,$   & 88   \\
      TPC 5/TPC 7 & $z$ & 2.08  & $\ \ 0.103 \pm 0.004$ & 2612 \\
      \midrule
      TPC 1/(3)/TPC 7 & $z$ & 4.16  & $-0.08 \pm 0.04$ \\
      TPC 1/(5)/TPC 7 & $z$ & 4.16  & $\ \ \, 0.23 \pm 0.01$  \\
      \bottomrule
    \end{tabular}                                                                         
\end{table}

The method demonstrated here has direct implications for similar
studies using the full DUNE FD.  All the gaps between the
APAs, both in the drift and $z$ directions, will need to be understood
for accurate reconstruction and are essential in order to make the
precise physics measurements with DUNE.  For example, the estimation
of the momentum of exiting muons using multiple scattering requires
precise understanding of the relative alignment of detector 
components~\cite{Antonello:2016niy,Abratenko:2017nki}, and the reconstruction
of the energies of showers crossing TPC boundaries is sensitive to the sizes
of the gaps.

%% file: apacrosser_section.tex
\section{Measurement of {\boldmath $t_0$} from tracks crossing the anode planes}
\label{sec:apagap}

The 35-ton prototype collected data from tracks that pass from one drift volume
to the other, thereby passing through the APA planes.
The 35-ton is the only planned experiment in the LAr prototyping programme 
in which the APAs read out drifting charge on both sides simultaneously, a feature of DUNE's Far Detector.
The ProtoDUNE-SP prototype also read out two-sided APAs, but there is no
drift field on the cryostat side of each APA, so deposited
charge does not drift towards the APA~\cite{protodunesp}.  
However, charge deposited between its wire planes could drift to both sides of the ProtoDUNE-SP APA.

Since these tracks in the 35-ton prototype cross the planes, it is possible to measure the
arrival time $t_0^{\rm{TPC}}$ of the cosmic ray by requiring that the two track segments are aligned across
the anode planes.  
An incorrect $t_0$ would introduce a common timing
offset and have the effect of moving both track segments
closer to or further from the APAs.  The value of $t_0^{\rm{TPC}}$ can then be compared to 
that measured by the CRCs, $t_0$, which is the event
trigger time.  Two such tracks are visible in figure~\ref{fig:evd_crossing}.

Tracks with a shallow APA-crossing angle are selected, to ensure
sufficient hits in each drift region.  
Only collection-plane hits within the triggered counter shadow are used.
A linear least-squares fit is
applied to the track segment in each drift region separately, and $t_0^{\rm{TPC}}$
is determined by aligning the two track segments.  
The method was tested with simulated data and used on the detector data to determine
the relation between $t_0$ and $t_0^{\rm{TPC}}$.    The approximate resolution on the
timing difference $t_0^{\rm{TPC}}-t_0$ is $\pm 1\mu$s per track in simulated events, and 
$\pm 3\mu$s per track in the data.
A systematic offset between $t_0$ and $t_0^{\rm{TPC}}$ of 62 ticks ($31 \mu$s) 
is observed. Possible sources of the delay include the buffering in the front-end electronics, triggering,
and event readout.
More details of the method are provided in Reference~\cite{Wallbank:2018otb}.

\begin{figure}
  \centering \includegraphics[width=14cm]{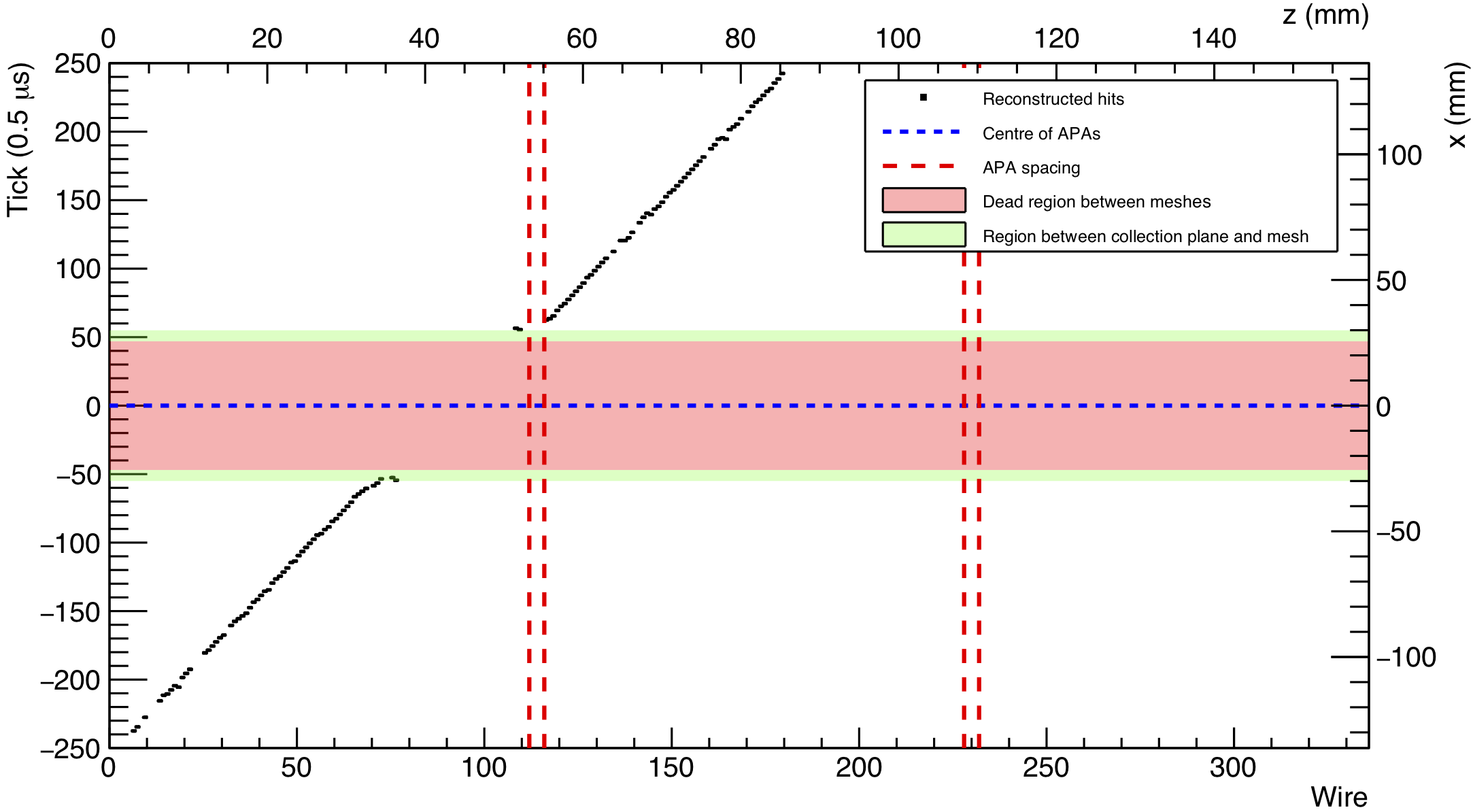}
  \caption{Reconstructed hits from an APA crossing track deposited
    near the APAs. Data displayed negative times is from the shorter
    drift volume. A ``hook''-like effect due to backwards-drifting charge is
    evident.}
  \label{fig:apa_crosser}
\end{figure}

Further studies using the APA crossing tracks involved studying the
distributions of the readout time of each hit associated with the
crossing track, relative to $t_0^{\rm{TPC}}$.  There is a sharp peak in this
distribution corresponding to the arrival time of the cosmic ray~\cite{Wallbank:2018otb}.
However, this peak was not present for hits on tracks which cross
the short center APA.  This APA is the only one without a grounded
mesh.   Hits populating the peak at the
cosmic-ray arrival time are thus ascribed to charge deposited between the collection-plane
wires and the grounded mesh.  This charge drifts in an opposite direction with respect
to charge drifting from the bulk of the TPC.
Figure~\ref{fig:apa_crosser} shows hits from an APA-crossing track
in the time vs. wire plane.  Such tracks exhibit hook-like features in the event
displays as the backwards-drifting charge arrives on the collection wires at
positive drift times just as forward-drifting charge.

The grounded mesh provides a uniform ground plane over the face of each APA in which
it is installed.  In APAs with grounded meshes,
the distribution of hit times is the same for wires passing over
the center of the APA as it is for wires passing over the grounded frames, indicating
that the meshes are performing as designed.

%% file: lifetime_section_v5.tex
\section{Electron lifetime measurement}\label{sec:lifetime}

Free electrons in the LAr attach to electronegative impurities, such
as oxygen and water, reducing their drift velocity.
The attached charge is still collected at the anode,
just much more slowly and at much later times than the unattached charge, and thus
it does not contribute to signal pulses.
The electron lifetime, $\tau$, is defined by the exponential
decay of the charge measured at the anode, $Q_{\rm{meas}}$, with drift
time, $t$,
\begin{equation}\label{eq:lifetime}
Q_{\rm{meas}} = Q_{0} e^{-t/\tau},
\end{equation}
where $Q_{0}$ is the charge liberated in the ionization after
recombination.  A measurement of the lifetime is necessary in order to
correct the measured charge for each hit in each event which is needed for
energy reconstruction and particle identification.

As mentioned in section \ref{sec:detector} dedicated purity monitors were
used for online measurements.
As measured by purity monitor~\#2, the mean lifetime in the cryostat but
outside of the TPC for the five-day dataset used below is
$2.8\pm0.1$~(stat.)~$\pm$~1.1~(syst.)~ms, as shown in figure~\ref{fig:PrM2}.
The purity was observed to fluctuate during this period by about 4\%.

\subsection{Electron lifetime analysis}
In addition to the dedicated purity monitors, the electron lifetime was measured offline with the reconstructed cosmic-ray muon tracks in the active volume of the TPC.  
The electron lifetimes in several liquid argon TPCs have been measured with
tracks using methods similar to the one described here
\cite{lifetimeT600, lifetime120L,
  lifetimeT600_again,lifetime-argoneut,lifetime-longbo}.  Additional details on the method
described here can be found in~\cite{Thiesse:2017qom}.

In this analysis, hits found with the THB and associated with reconstructed
tracks (section~\ref{sec:hitstracks}) are used to
determine the lifetime. 
Distributions of the hit $dQ/dx$ values are formed in 22 regions of drift
time, corresponding to drift regions $\approx$10~cm across in the long drift volume.
Each of these distributions is fit to a
Landau distribution convoluted with a Gaussian representing the detector
response.   

The data used in this analysis consist of 17,490 events, triggered on
east-west crossing muons, from five consecutive days of the Phase~II
run when the cathode HV was stable, the purity monitors reported greater than
2~ms lifetime, and the detector was in the low-noise
state. Fluctuations in the electron lifetime over the course of the
five-day period are not studied in this analysis. 

\begin{figure}
\centering \includegraphics[width=0.7\textwidth]{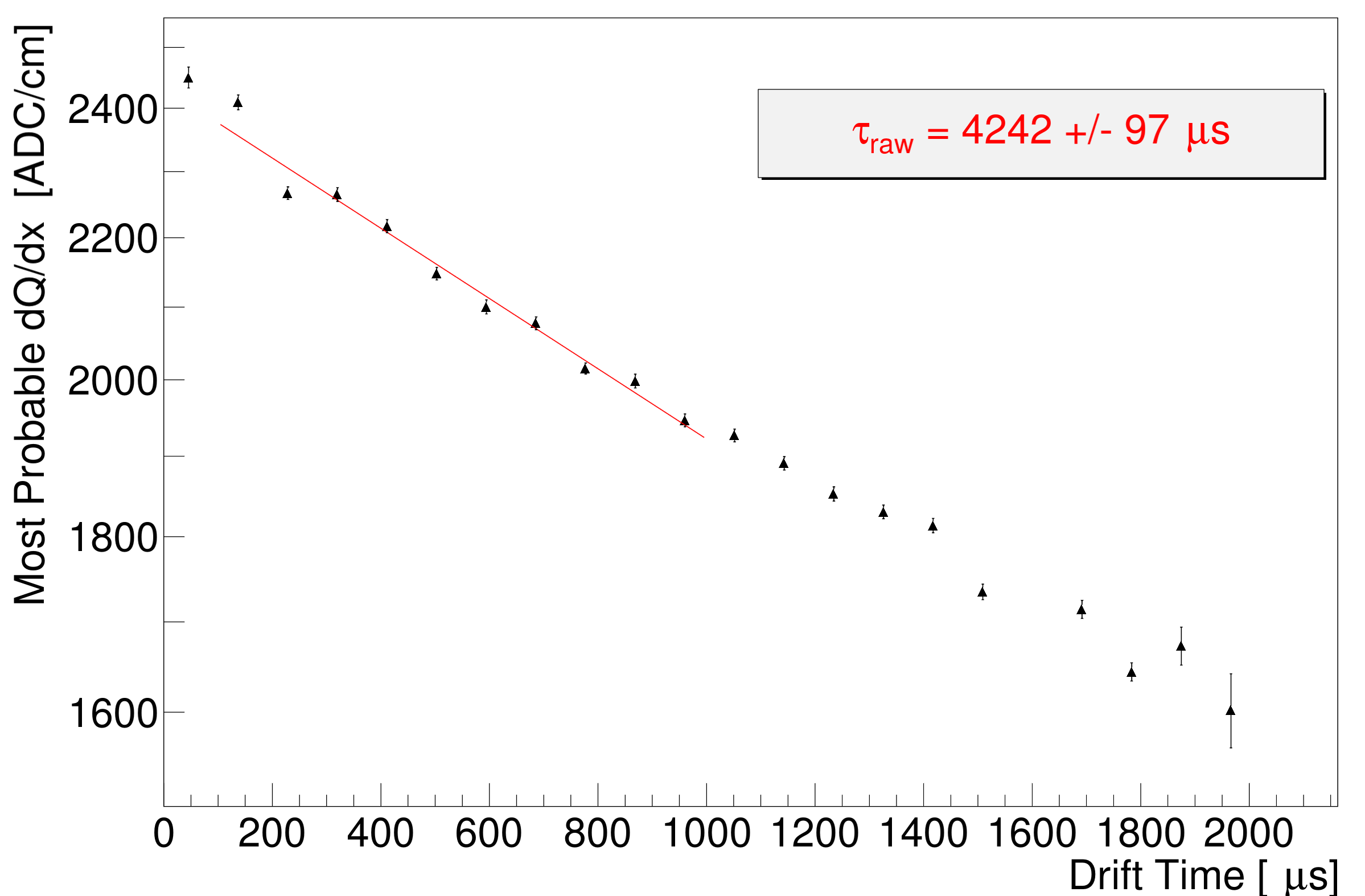}
\caption{Most probable hit $dQ/dx$ measured at the anode, as a
  function of drift time. An exponential fit to the data in the fiducial range is shown in
  red.}
\label{fig:landMPV}
\end{figure}

For this dataset, the fitted most probable value (MPV) of $dQ/dx$ as a function of drift
time in the TPC is shown in figure~\ref{fig:landMPV}.  A fit to
a decreasing exponential yields an
observed raw lifetime of
$\tau_{\rm{raw}}=4.24\pm0.10$~(stat.)~ms. 
Only drift times from 100~$\mu$s to 1000~$\mu$s are included in the fit
in order to reduce the impact of biased ionization MPV measurements~\cite{Thiesse:2017qom}. 
The bias corrections and systematic uncertainties are described below.

\subsection{Simulating the lifetime measurement bias}\label{sec:lifetimebias}

The S/N ratio and the particular electronic noise
characteristics of the 35-ton create biases in the electron lifetime
measurement.   These biases arise from the fact that the hit-finding efficiency
is a strong function of the hit charge, with low-charge hits being the most difficult
to detect.  Charge resolution and contamination from noise hits contribute as well.

In order to evaluate the bias in the raw electron lifetime measurement $\tau_{\rm{raw}}$,
simulated samples with known lifetimes are analyzed in the same way as the
data and the lifetimes measured in the simulated samples are compared with that
measured in the data to invert the bias.
Because the Monte Carlo simulation does not
replicate the changing noise amplitudes and spectra observed in the data,
nor noise coherence between channels, the data itself is used as the
noise model.  Cosmic-ray signals are simulated with the CRY~\cite{CRY} event
generator and the LArSoft toolkit~\cite{larsoft} which uses
GEANT4~\cite{Agostinelli:2002hh,Allison:2006ve,Allison:2016lfl}
 as the physics simulation package.  The CRY event generator is 
 configured for this analysis to produce a single muon per triggered readout
 with momentum and direction sampled from a realistic parameterisation of 
 the cosmic-ray muon flux at the Fermilab site.  The simulated
events are produced with no simulated noise.  Raw digits thus simulated are then added
to data raw digits, selected sufficiently far away in time from triggered
cosmic rays to eliminate trigger bias.  Untriggered cosmic rays form a component
of the background and are present in the data used as the background model.

While the noise is modeled with data, the amplitude of the signal is a parameter
input to the simulation and is therefore a source of systematic uncertainty.  Samples of Monte Carlo overlaid
with data were made with signal scalings varying by a factor of four, and the resulting
$dQ/dx$ distributions compared with the data in order to constrain the signal scaling
and its uncertainty, which is approximately 15\%.  The corrected lifetime, obtained by
interpolating the simulated lifetime measurements as a function of input lifetime, is
$4.12\pm 0.17$~(stat)~ms.

\subsection{Systematic uncertainties}\label{sec:lifetimesystematics}

The systematic uncertainty associated with the biases introduced by
the noise is taken as the magnitude of the bias shift in the lifetime
calculated in the previous section, $4.24-4.12=0.12$~ms, or 2.9\%.
This includes the effects of low hit finding efficiency on
the true Landau MPV and the poor charge resolution for the relevant
region of hit charge, both caused by the high level of noise in the
detector.

Another source of systematic uncertainty is due to the accumulation of positive
space charge in the TPC, which, because of their low mobility in
comparison to the negative drift electrons, distorts the electric
field~\cite{palestini}.  The field distortion impacts
the recombination fraction~\cite{argoneut-recombination}
as a function of drift distance, which can mimic the effect of electron lifetime.
The fractional systematic uncertainty on the lifetime
due to this source is estimated to be 7.8\%.

Uncertainties due to the effects of transverse diffusion, channel-to-channel
gain variations and signal modeling errors in the Monte Carlo simulation are estimated
to contribute a 5\% fractional systematic uncertainty on the lifetime measurement ~\cite{Thiesse:2017qom,ublifetime}.

The measured lifetime of 4.12 $\pm$ 0.17~(stat) $\pm$ 0.40~(syst)~ms
is consistent with the average of the purity monitor
measurements, 2.8 $\pm$ 0.1~(stat) $\pm$ 1.1~(syst)~ms, over the same
span of runs.  The systematic uncertainty on the purity monitor measurements
is assessed from variations seen in the purity measurements when the operating
voltages were changed and uncertainties in the measured signal peak heights and voltages
of the anodes and cathodes in the purity monitors.  The largest part, however, is
estimated from the vertical stratification observed in the measurements mentioned
in section~\ref{sec:run}, and thus is not an uncertainty on the purity monitor measurements
of the electron lifetime of the liquid argon near the monitors, but rather it is an uncertainty
on the use of those measurements to estimate the electron lifetime averaged over the TPC
volume used in the measurement presented in this section.

\begin{comment}

\end{comment}

%% file: diffusion_section.tex
\section{Event time determination from pulse properties}
\label{sec:diffusion}
Measurement of the electron diffusion constants was one goal of the 35-ton analysis.
However, as mentioned in previous sections, the observed high noise levels led to
poor charge resolution. 
A precise measurement of
the longitudinal and transverse constants of diffusion is therefore
not possible.
Instead, a novel method of interaction time
determination using the effects of longitudinal diffusion and charge attenuation
due to electron lifetime has been developed and is presented below. A complete description of the
method is provided in Ref.~\cite{Warburton:2017ixr}.

The mechanism by which electron diffusion in liquid argon occurs is
discussed in
Refs.~\cite{AtrazhevTimoshkin,shibamura,derenzo1,derenzo2} and
early measurements are given. A set of
recent measurements for electric fields between 100 and 2000~V/cm
is presented in Ref.~\cite{Li:2015rqa}. The diffusion of electrons is
not isotropic. The component transverse to the drift field, called
transverse diffusion, and the component parallel to the drift field,
called longitudinal diffusion, are normally measured
separately. Longitudinal diffusion is generally smaller than
transverse diffusion. Longitudinal diffusion has the effect
of broadening the distribution of arrival times of the electrons at
the anode plane, while transverse diffusion distributes electrons
among neighbouring wires on the anode plane.  The effects of transverse
diffusion are more difficult to measure, as hits on neighbouring wires occur
at similar times, and the net effect is a worsening of the charge resolution,
which is also impacted by the detector noise.

Hits and tracks are reconstructed using the GHF and PMA respectively,
which are described in section~\ref{sec:hitstracks}.  The width
$W$ of a hit on a wire is defined to be the standard deviation of the
Gaussian function fit to the filtered ADC waveform as a function of time.
Longitudinal diffusion causes the average value of $W$ to increase
with drift distance.  The integrated charge of a hit is denoted $Q$.
$W$ and $Q$ both depend on the angle of the track with
respect to the electric field, and the distance from the wire
plane. The hit charge $Q$ is further sensitive to the electron
lifetime of the drifting medium.  The ratio $R=W/Q$
is less dependent on the track angle than either $W$ or $Q$, but it
contains distance sensitivity from both $W$ and $Q$.

This analysis uses tracks associated with East-West CRC coincidence triggers,
which are described in section~\ref{sec:ewcounters}. 
Cosmic rays which give rise to these triggers
consist predominantly of minimum-ionizing muon tracks that cross many
collection-plane wires.  A range of drift distances is covered
by using different counter pairs as triggers.  Only the collection-plane
wire signals are used because of the larger S/N ratio.  Data from noisy
wires are excluded, and $\delta$-rays are identified and excluded.
The reference track times ($t_0$) are obtained from the counter coincidence trigger time, and the
reference track positions are computed from the difference between the hit times and the
reference times, multiplied by the drift velocity.  

%\subsection{Extracting Interaction Times from Hit Properties}

%A set of lookup tables is constructed, which map the reference distance (or time)
%to the average values of $W$ or $W/Q$, in 10~cm wide bins of the reference distance.
The averages of the distributions of the variables $W$ and $W/Q$ are computed as functions of the reference distance in 10 cm bins. 
In the case of $W$, the data are also binned in track angle.  Linear fits to
these functions are used in order to parameterise and invert the relationship between the discriminant
variables and the distance.  Each hit's estimated distance is obtained from the
linear parameterisation, and the estimated distances are converted to interaction times.
The estimated interaction time for a track $t_{\rm{int}}$ is then the average of the times for each hit.
The distributions of $W$ and $W/Q$ are broad, and thus 100 hits are required in order to
estimate the interaction time.

\begin{figure}[ht]
  \centering
  \begin{subfigure}{0.49\textwidth}
    \centering
% old, with MC overlaid
%    \includegraphics[width=\textwidth]{Overlay_AvTimeDiff_RMS}
    \includegraphics[width=\textwidth]{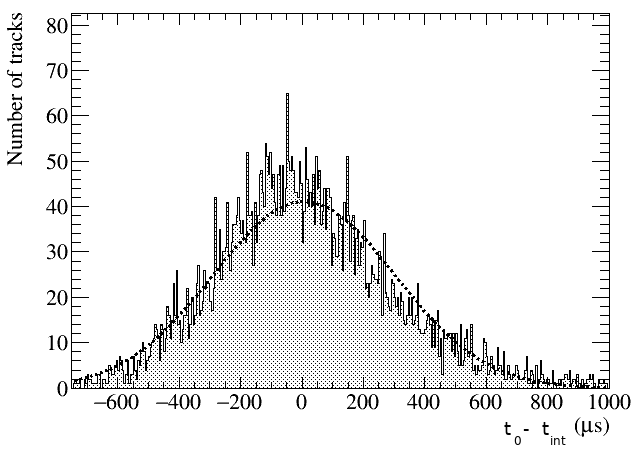}
    \caption{   }
    \label{fig:DiffOverlayAvDiff_RMS_T}
  \end{subfigure}
  \begin{subfigure}{0.49\textwidth}
    \centering
% old, with MC overlaid
%    \includegraphics[width=\textwidth]{Overlay_AvTimeDiff_RMS_Int}
    \includegraphics[width=\textwidth]{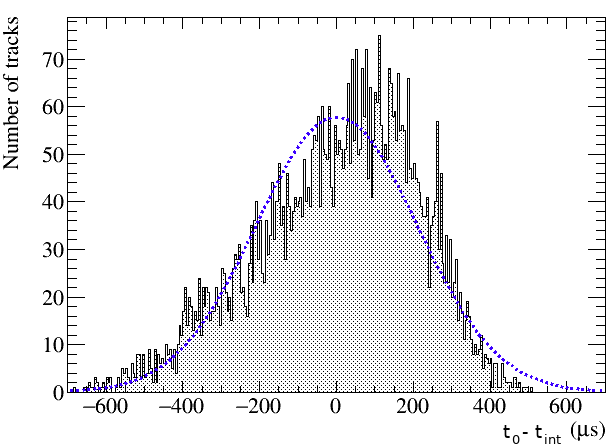}
    \caption{ }
    \label{fig:DiffOverlayAvDiff_RMSInt_T}
  \end{subfigure}
  \caption{
     The distributions of the difference between the trigger time and the interaction time estimated from
     hit properties in 35-ton prototype data.  Panel (a) shows the
     distribution using the $W$ metric, and panel (b) shows the distribution using the $W/Q$
     metric. Gaussian functions are fit to the distributions.  A bias of 240~$\mu$s has been subtracted in (a) and a bias of 171~$\mu$s has
     been subtracted in (b).
}
  \label{fig:DiffOverlayAvDiff}
\end{figure}

The 35-ton prototype data are used to estimate the accuracy and the bias in the interaction time
reconstructed from the hit charges and widths, by comparing the reconstructed interaction times with
the trigger times. The distributions of the time differences are shown in
figure~\ref{fig:DiffOverlayAvDiff} for the $W$ and $W/Q$ discriminant variables.
Biases of 240~$\mu$s and 171~$\mu$s are observed in the $W$ and the $W/Q$ analyses.  Both biases have been subtracted from the distributions shown in figure~\ref{fig:DiffOverlayAvDiff}.

When using the $W/Q$ metric, the FWHM of a
Gaussian fitted to the distribution is 210 $\mu$s for the 35-ton prototype
data set. This is much less than the nominal drift
time of 5200 $\mu$s in the 35-ton prototype at a drift field of 250~V/cm. 
The result of this is that it should be possible to
separate out tracks across a drift volume, using just the effects of
longitudinal diffusion and hit charge. The accuracy to which this can be done is
still not good enough to replace determinations using external sources
such as counter coincidences, or flashes of scintillation
light.   In some events, multiple cosmic ray particles may arrive at different times
and locations, introducing ambiguity in the association between flashes and charge.
%In the DUNE Far Detector, low-energy events, for example from %supernova burst interactions,
%may not produce enough light to be detected by the photon detectors.  
In such cases,
using hit parameters will be useful in determining the distance of an interaction to the anode plane. More details of this analysis can be found in Ref.~\cite{Warburton:2017ixr}.

%% file: summary_section.tex
\section{Summary}
\label{sec:summary}

The 35-ton prototype successfully demonstrated in Phase~I that liquid
argon of sufficient purity could be maintained in a membrane cryostat
with adequate filtering and circulation.  Phase~II confirmed that this
is also the case when a time-projection chamber and associated
electronics and cabling were installed.  The Far Detector design
evolved after the 35-ton design was finalised, and  the noise
characteristics of the 35-ton prototype detector made analyses challenging.
Nonetheless, a number of analyses of the cosmic-ray data are possible and are presented
here:  the relative alignment of the TPC and the external counters using
cosmic-ray muons, the relative alignment
of the anode plane assemblies, the timing offsets between the TPC and the
trigger, the electron lifetime, and
a novel method of constraining the interaction time from charge and hit width.
These analyses study the unique features of a modular liquid argon TPC similar
to that proposed for the DUNE single-phase Far Detector modules.

\section{Acknowledgments}
This material is based upon work supported in part by the following: the U.S. Department of Energy, Office of Science, Offices of High Energy Physics and Nuclear Physics; the U.S. National Science Foundation; the Science and Technology Facilities Council of the United Kingdom, including Grant Ref: ST/M002667/1; and the CNPq of Brazil. Fermilab is operated by Fermi Research Alliance, LLC under Contract No. DE-AC02-07CH11359 with the United States Department of Energy. We would like to thank the Fermilab technical staff for their excellent support.

%% file: bibliography.tex
\bibliographystyle{model1-num-names}
%\bibliography{sample.bib}